# Konjac Bio-Molecules Assisted – Rod / Spherical shaped Lead Nano Powder Synthesized by Electrolytic Process and Its Characterization Studies

T. Theivasanthi * and M. Alagar

*Centre for Research and Post Graduate Department of Physics, Ayya Nada Janaki Ammal College, Sivakasi - 626124, Tamilnadu, India.*

______________________________________________________________________

**Abstract:** *Synthesis and structural characterization of Pb nanoparticles by electrolysis using a bioactive compound - konjac aqueous extract is the main aim of this study. This method is a unique, novel, low cost and double-step procedure with good reproducibility and has not been used for nanoparticles preparation so far. Konjac extract has been added to prevent the oxidation of Pb nanoparticles. Also the synthesized nanoparticles have been dried in open air to observe their stability. Various types of characterization tools like XRD, SEM, Particle Size Analyzer, TEM-EDS, DSC, AAS and FT-IR have been utilized to study characters of the end product. Anti-bacterial Studies has also been done. After completion of synthesis process that we have made an attempt to change the shape of the synthesized nanoparticles by the influence of sunbeams and to find the effects of the sunlight on nanomaterials.*

______________________________________________________________________

## 1. Introduction

In the past decade, one-dimensional nanostructures, such as nanotubes and nanowires, have attracted extraordinary attention for their novel physical properties and potential applications in constructing nanoscale electronic and optoelectronic devices.[1–2]. The fabrications of nano sized materials are of great scientific and technological importance in applications such as photonic materials, microchip reactor, miniaturized sensor, separation technologies, and non-linear optical apparatus [3-4]. Nanoparticles are being synthesized in various methods including electrolysis and many irradiation methods such as sonochemical [5], UV or vis photoirradiation [6], laser pulse [7] and γ-irradiation methods [8] and so forth.

Nanomaterials are usually characterized for their chemical, physical and optical properties in their pure form or in simple idealized matrices. However, when nanomaterials interact with biological systems, their properties are changed significantly, affecting their functionality and behavior. While, on contact with a biological fluid they may become coated with proteins and other bio-molecules. The potential interactions of nanoparticles with biological systems may also be desirable characters in sometimes.

______________________________________________________________________

**\*Corresponding author:** Phone: +91-9245175532
 *E-mail*:  theivasanthi@pacrpoly.org

Investigating the formation mechanism of nanoparticles is always important for the fabrication of advanced integrated functionalized-materials based on the nanoparticles. As more is understood about the properties and the mechanism of formation of these nanoparticles, a better control of their size, shape and applications can be achieved. Solvent, pH value, photo-sensitizer, the concentration and category of stabilizer, irradiation intensity, wavelength, and irradiation time are all important factors that must be considered for shape and size controlled synthesis of metal nanoparticles. [9]

Size and shape provides important control over many of the physical properties (viz., melting point, magnetism, specific heat, conductivity, band gap, etc.), luminescence and optical properties, chemical properties and catalytic properties of nanomaterials [10]. Up till now, scientists have controlled particle size and shape for several metal and semiconductor particles and the controlled size/shape are very useful in various applications like SERS[11], catalysis [12]. Common methods for size control employ capping agents [13], such as surfactants ligands, polymers or dendrimers, to restrict the growth in the nanometer regime.

Capping agents assisted synthesis methods usually produce spherical particles due to the low surface energy associated with such particles. The procedures for shape-controlled preparation of monodispersed nanoparticles, on the other hand, are limited. UV photoactivation technique has been used to achieve controlled nucleation for the regulation of particle size in monodispersed gold solution. In photochemical reduction, hydrated electrons or free organic radicals formed by photo-irradiation (e.g., by UV or visible light) reduces the metal ions to metals. The photolysis of phenols in aqueous solution produces hydrated electrons. Photo-initiators like phenolic compounds absorb light in the UV–visible region and form reactive intermediates such as free radicals. [14]. Methanol is also a photoinitiator, used for such purpose [15]. These intermediates (or hydrated electrons) are serving as a reducing agent and prevents the oxidation of lead nanoparticles.

Miranda and Ahmadi [16] have studied the growth of gold nanorods by photochemical reduction of HAuCl4 in micelle solution. They found that the anisotropic growth of gold nanoparticles was intensively affected by the wavelength of UV light source and their photon flux. Esumi et al. [17], have prepared rod like gold particles by UV irradiation and reported silver, gold and copper strongly absorb light in the visible region due to surface plasma resonance. Also, they have elucidated in another report [18] that the number of surface amino group of dendrimers was a key factor to control the particles with UV irradiation method.

Metals and non-metals present on earth are in constant association with biological components. This interaction of metals and microbes are natural and continuous since the inception of life provides the exciting areas of research. Stabilization of metal nanoparticles by protein is possible one [19]. It is reported earlier that proteins can bind to nanoparticles either through free amine groups or cysteine residues in the proteins [20] and via the electrostatic attraction of negatively charged carboxylate groups in enzymes present in the cell wall of bacteria [21].

*Amorphophallus campanulatus (Roxb.) Blume. ex Decne* (Synonym : *Amorphophallus paeoniifolius; Araceae)* commonly known as Konjac is a tuberous, stout, indigenous herb, 1.0-

1.5 m in height. The tubers contain an active enzyme amylase, amino acids, carbohydrates, saponin. Antioxidants are chemical substances that donate electron. The aqueous extract of *A.campanulatus* tuber has more significant antioxidant activity and more reducing power as compared to methanol extract [22]. 46.33 mg g$^{-1}$ of flavonoidal content, 12.67 mg g$^{-1}$ of phenolic contents and tannins are present in the air dried powder of *A.campanulatus* tubers [23].

Konjac has listed as one of the top ten healthiest foods by World Health Organization (WHO). It is highly nutritious, with over 45% of Glucomannan, 9.7% of crude protein, 16 different kinds of amino acids up to 7.8%, 7 essential amino acids up to 2.5% , low fat and high in fiber. Glucose fructose and sucrose are also present in konjac. Glucomannan is a water-soluble dietary fiber and it is a polysaccharide (hemicelluloses) straight-chain polymer, with a small amount of branching. It is a high-molecular compound and a super absorbent; its size expands by 80-100xs after absorbing water. 30% to 40% of total production of flying powder is produced from konjac tuber. The Konjac flying powder is a polyol (natural polymer). It is used in precipitation of soluble heavy metal ions in wastewater.

Lead nanostructures are attractive materials for its potential applications in lead batteries, catalysis, superconductor [24] and photonic crystal [25]. Earlier study reports that hexagon and flower shaped lead oxide ($PbO_2$) are the end product in electrolysis process (anodic oxidation of lead) using lead sheet as anode and distilled water (with very low ionic conductivity _ 6-10 μS/m and pH=6.5) as electrolyte [26].

It is known that Pb nanoparticles are highly reactive and transfer electron to $N_2O$, $O_2$, etc. [27]. The Pb hollow nanoparticles were produced only when a 1:3 ratio of the Pb precursor to $NaBH_4$. Black colour Pb hollow nanoparticles gradually changed from black to pale yellow on exposure to air. The SAED, XRD, and HRTEM data of these pale yellow particles indicated that a mixture of Pb, litharge (tetragonal PbO), massicot (orthorhombic PbO), and $Pb_5O_8$ formed. When a $Pb(OAc)_2·3H_2O$ aqueous solution instead of a $Pb(OAc)_2·3H_2O$-PVP solution was added to an aqueous solution of $NaBH_4$ and PVP, only solid nanoparticles (5 ± 1.8 nm) were formed, and their hollow counterparts were not observed [28].

In this study, we will present preparation and structural characterization of Pb nanoparticles synthesized by electrolysis using a bioactive compound - konjac aqueous extract as stabilizing agent to prevent from oxidation. Our method is a unique, novel and double-step procedure with good reproducibility. To our knowledge, such a procedure has so far not been used for nanoparticles preparation. Also, we have made an attempt to find the effects of the sunlight on nanomaterials and got morphologically changed - rod like Pb nanoparticles.

It explicates that the sunlight can be utilized for dual function i.e. as a dryer for the synthesized materials and as a morphological changer. We found the shape of the konjac extract added Pb nanoparticles changed to rod shape by the influence of sunbeams. It is pertinent to note that we have made these morphological changes after completion of the entire synthesis process. Avoiding oxidation is the main challengeable task while on metal nanoparticles preparation particularly in Pb nanoparticles synthesis. Drying Pb nanoparticles in open air and sunlight indicates that this task is defeated / achieved by adding konjac extract.

## 2. Experimental Method

### 2.1. Chemical Effects of Current

The phenomenon of electrolysis is an important chemical effect of electric current. It passes and conducts electric current through lead nitrate solution with the resulting chemical changes. It is the dissociation process of lead nitrate solution and the dissociated ions that appear at the two electrodes. Lead nitrate solution is conducting electricity due to drifting of ions (positive cations and negative anions).

Solid $Pb(NO_3)_2$ is made up of $Pb^{2+}$ and $NO_3^-$ ions bound by a strong force of attraction. The thermal energy at room temperature is only 0.03 eV per molecule which is not enough to dissociate $Pb(NO_3)_2$ into $Pb^{2+}$ and $NO_3^-$ ions. When $Pb(NO_3)_2$ is dissolved in water, the force of attraction is greatly reduced because of the high dielectric constant (= 81) of water. In fact, the force reduces by a factor of 81, and the thermal energy is sufficient for the ionization process that is to dissociate $Pb(NO_3)_2$ completely into $Pb^{2+}$ and $NO_3^-$ ions.

### 2.2. Electrolytic Conduction Phenomenon

When a steady current flow into the electrolytic cell, due to ionization process, the $Pb(NO_3)_2$ solution is dissociated, lead is removed from the anode and deposited on the cathode. Electrons flow from the positive terminal to the cathode, $Pb^{2+}$ ions move towards cathode and $NO_3^-$ ions move towards anode. Oxidation reaction takes place at the anode and reduction reaction occurs at the cathode, and Pb atoms get deposited at the cathode. Lead has valency two and two electrons circulate for the deposition of one lead atom.

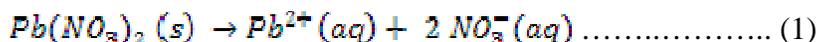
$$Pb(NO_3)_2\ (s) \rightarrow Pb^{2+}(aq) + 2\ NO_3^-(aq) \quad \ldots\ldots\ldots\ldots\ldots (1)$$

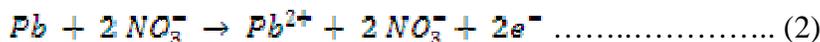
$$Pb + 2\ NO_3^- \rightarrow Pb^{2+} + 2\ NO_3^- + 2e^- \quad \ldots\ldots\ldots\ldots\ldots (2)$$

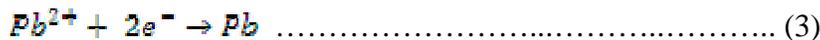
$$Pb^{2+} + 2e^- \rightarrow Pb \quad \ldots\ldots\ldots\ldots\ldots\ldots\ldots\ldots\ldots\ldots (3)$$

### 2.3. Electrolysis Synthesis

An efficient fabrication method for pure lead nanoparticles supported on Konjac extract has been developed using electrolysis method. This method has produced lead nanoparticles with non-oxidized surfaces and the konjac extract provides excellent chemical stability. The experiment was performed in an electrochemical bath with two electrode set up. High purity lead rod was utilized as a working electrode (anode). A stainless steel rod was served as the counter electrode (cathode). The distance between the electrodes was kept 3 cm. The volume of the electrochemical bath was 4x3x3 $cm^3$.

5 gm of $Pb(NO_3)_2$ (Lead(II) Nitrate) salt was kept in a clean glass vessel. 100 ml of distilled water was poured in this vessel and stirred well until the lead nitrate salt was completely dissolved in the water (clear and colourless solution). This solution was transferred to the electrochemical bath and used as an electrolyte. A constant voltage of 15 volts was applied between the electrodes using a power supply for a time span of 10 minutes. This process is shown in Figure 1.

At the end of the process, deposition of Pb nanoparticles was observed and they were removed from the cathode and the Pb nanoparticles (settled down on the electrolytic cell) were also removed from the bath. For stabilizing these nanoparticles by konjac extract solution, they were kept aside. For making konjac aqueous extract, 20 gm of konjac tuber was sliced / cut into many pieces. These pieces were put in a vessel and a 100 ml of distilled water poured in it. This vessel was kept in a stove and boiled for 10 minutes. At the end, konjac pieces were separated and the konjac aqueous extract was decanted.

The electrolyte $Pb(NO_3)_2$ is an oxidizing agent. The Pb nanoparticles are higly reactive material and they can be easily oxidized. So, the stabilization of lead nanoparticles is essential. A few drops of Konjac extract (stabilizing agent) was added to the synthesized Pb nanoparticles and were kept in a hot air oven at 50 °C for two hours / until it dried. After completion of drying process, their appearance was very fine and powdery in nature. Their structural characterizations were studied and results confirmed the formation of lead nanopowder (with spherical shaped particles).

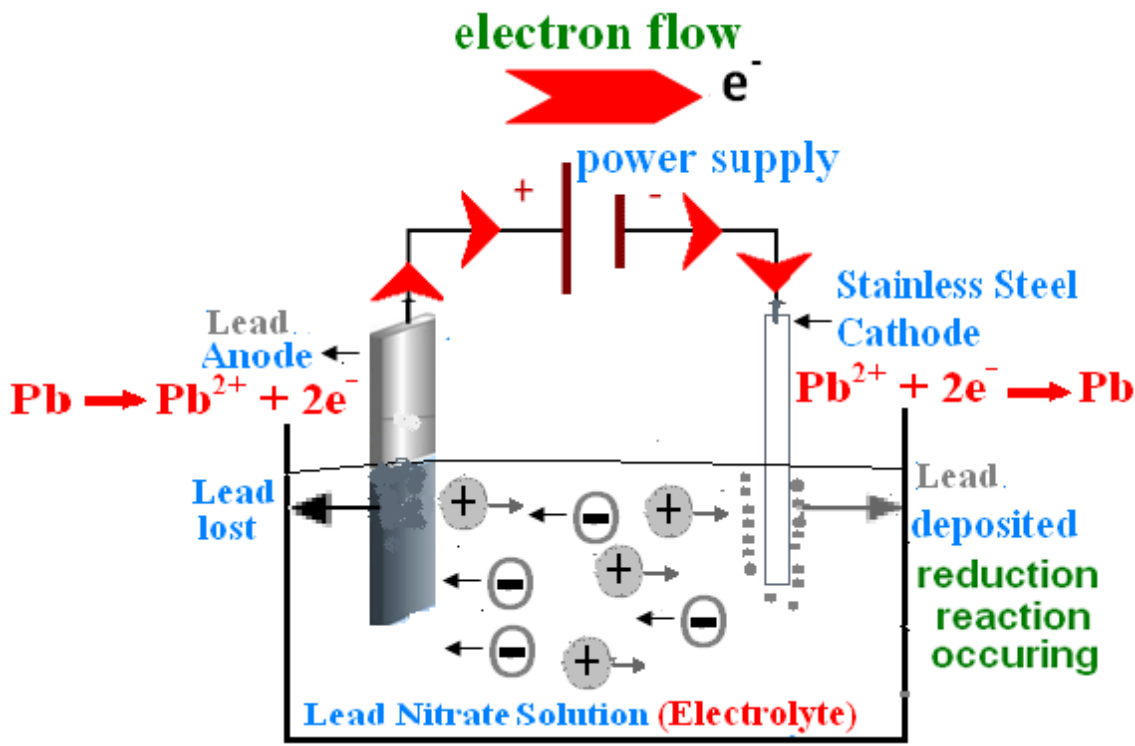

**Fig.1.** Schematic Diagram of $Pb(NO_3)_2$ Solution.

To get rod shaped Pb nanoparticles (to change the spherical shape to rod shape), the same synthesis procedures were adopted except small change in drying process. The Pb nanoparticles were dried in sunlight instead of drying in hot air oven (Model: HASTHAS oven with size 355 mm x 355 mm x 355 mm and maximum temperature 300 °C).

The XRD analysis of the prepared sample of Pb nanoparticles was done using a X'pert PRO of PANalytical diffractometer, Cu-Kα X-rays of wavelength (λ)=1.5406 Å and data was taken for the 2θ range of 10° to 80° with a step of 0.0170°. The surface morphology was analyzed by using SEM (JEOL Model JSM-6360). In the SEM an electron beam is focused into affine probe and subsequently scanned over a small rectangular area. The beam interacts with the sample and creates various signals (secondary electrons, internal currents, photon emission, etc.). All these signals can be appropriately detected. The particle size of the Pb nanoparticles was analyzed by Dynamic Light Scattering method using particle size analyzer (Malvern Zetasizer nanosizer). Particle size was arrived based on measuring the time dependant fluctuation of scattering of laser light by the nanoparticles. TEM coupled with EDS (Model-ZEISS) was used to take high resolution, brightfield, dark field images of particles, to analyze the outer surface and inner structures of the particles. Thermal studies of the sample were analysed by Diffrential Scanning Calorimetry (DSC). Concentration of the sample was analyzed by Atomic absorption spectrophotometer (model SHIMADZU-AA 6300). Functional groups were analyzed by SHIMADZU FT-IR spectrometer. The antibacterial activities of Pb nanoparticles were studied against *Escherichia coli* by Cup and Plate method. Standard Zone of Inhibition (ZOI) was measured and evaluated from this microbiology assay.

## 3. Results and Discussions

### 3.1. X-Ray Diffraction Studies - Peak Indexing – Pb Nanorods

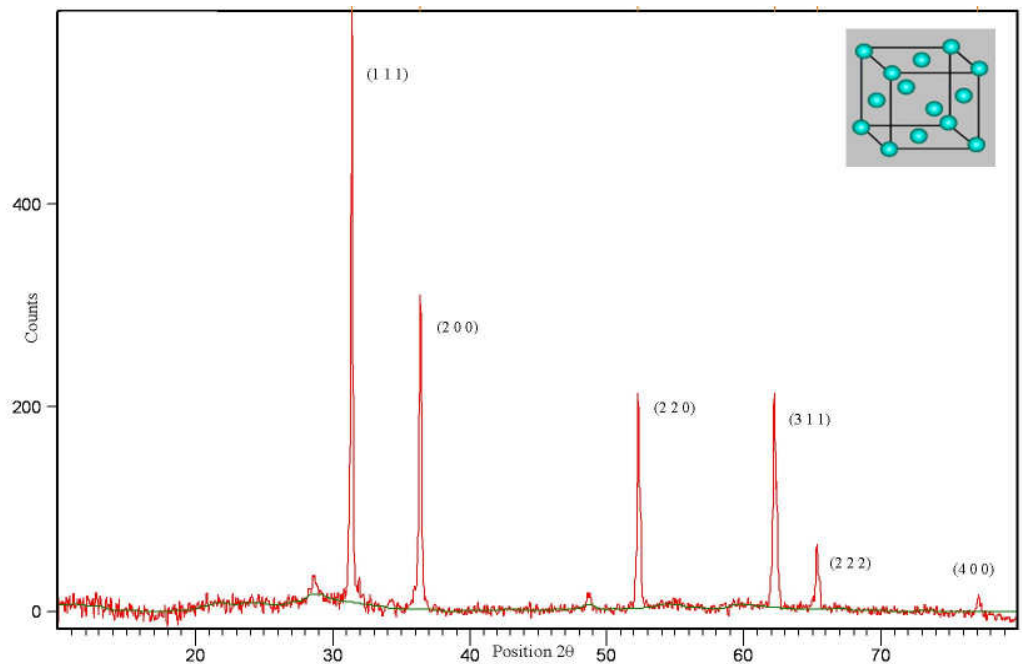

**Fig.2**. XRD showing Peak Indices & 2 θ Positions

The X-ray diffraction pattern of the lead nanorods synthesized by electrolysis method is shown in Figure 2 and the peak details are in Table.1. Indexing process of powder diffraction pattern is done and *Miller Indices* (hkl) to all peaks is assigned in first step. Indexing is done in two methods, same result is observed in both methods and the details are presented in Table.2 & Table.3. A dividing constant 24.229 (= 97.352 - 73.123) is found from the values in Table.2. and the integers from the approximately values of 1000 x $Sin^2 \theta$ / 24.229.

A number of strong Bragg reflections can be seen which correspond to the (111), (200), (220) (311), (222) and (400) reflections of Face Centered Cubic symmetry (FCC) lead. Absence of spurious diffractions indicates the crystallographic purity. [29]. There is no any spurious diffraction peak found in the sample. Each crystallographic facet contains energetically distinct sites based on atom density. High atom density facets such as (111) are known to be highly reactive [30]. The lead nanoparticles contain high atom density facets (111). The high intense peak for FCC materials is generally (111) reflection; h, k, l details all even or all odd number in each h, k, l; both of these facts are observed in the prepared sample.

**Table.1.** Intensity of XRD Peaks

| 2θ of peak (deg) | Height (counts) | FWHM (deg) | d-spacing (Å) | Relative Intensity (%) |
|---|---|---|---|---|
| 31.3777 | 574.62 | 0.1673 | 2.85096 | 100 |
| 36.3611 | 300.83 | 0.1840 | 2.47086 | 52.35 |
| 52.3101 | 205.37 | 0.1338 | 1.74894 | 35.74 |
| 62.2696 | 166.90 | 0.3346 | 1.49102 | 29.05 |
| 65.3477 | 56.00 | 0.3346 | 1.42804 | 9.75 |
| 77.0153 | 11.81 | 0.4080 | 1.23719 | 2.06 |

**Table 2.** Simple peak indexing.

| Peak Position 2θ | 1000 x $Sin^2 \theta$ | 1000 x $Sin^2 \theta$ / 24.229 | Reflection | Remarks |
|---|---|---|---|---|
| 31.3777 | 73.123 | 3 | (111) | $1^2+1^2+1^2 = 3$ |
| 36.3611 | 97.352 | 4 | (200) | $2^2+0^2+0^2 = 4$ |
| 52.3101 | 194.306 | 8 | (220) | $2^2+2^2+0^2 = 8$ |
| 62.2696 | 267.344 | 11 | (311) | $3^2+1^2+1^2 = 11$ |
| 65.3477 | 291.445 | 12 | (222) | $2^2+2^2+2^2 = 12$ |
| 77.0153 | 387.655 | 16 | (400) | $4^2+0^2+0^2 = 16$ |

**Table.3.** Peak indexing from d – spacing

| 2θ | d (Å) | $1000/d^2$ | $(1000/d^2)$ / 40.766 | hkl |
|---|---|---|---|---|
| 31.3777 | 2.85096 | 123.03 | 3 | (111) |
| 36.3611 | 2.47086 | 163.796 | 4 | (200) |
| 52.3101 | 1.74894 | 326.927 | 8 | (220) |
| 62.2696 | 1.49102 | 449.814 | 11 | (311) |
| 65.3477 | 1.42804 | 490.365 | 12 | (222) |
| 77.0153 | 1.23719 | 653.322 | 16 | (400) |

The remarkably intensive diffraction peak noticed at 2θ value of 31.3777 from the {111} lattice plane unequivocally indicates that their basal plane, i.e., the top crystal plane, should be the {111} plane. It has been suggested that this plane may possess lowest surface tension [31]. The intensity of peaks reflects the degree of crystallinity and broadening in diffraction peaks indicates very small size of crystallite [32]. The intensity of peaks in the Pb nanoparticles XRD reflects that the formed lead nanoparticles are crystalline and broad diffraction peaks indicate very small size crystallite. The size of the Pb nanoparticles estimated from Debye–Scherrer formula (Instrumental broadening) is 8 nm.

**Table.4.** Experimental and standard diffraction angles of Lead specimen

| Experimental | | Standard Lead – JCPDS: 65-2873 | |
|---|---|---|---|
| diffraction angle [2θ in degrees] | d.spacing Å | diffraction angle [2θ in degrees] | d-spacing Å |
| 31.3777 | 2.85096 | 31.2867 | 2.8574 |
| 36.3611 | 2.47086 | 36.2828 | 2.4746 |
| 52.3101 | 1.74894 | 52.2506 | 1.7498 |
| 62.2696 | 1.49102 | 62.1757 | 1.4922 |
| 65.3477 | 1.42804 | 65.2722 | 1.4287 |
| 77.0153 | 1.23719 | 77.0311 | 1.2373 |

**Table.5.** Ratio between the intensities of the diffraction peaks

| Diffraction Peaks | Sample Value | Conventional Value |
|---|---|---|
| (200) and (111) | 0.523 | 0.479 |
| (220) and (111) | 0.357 | 0.310 |
| (311) and (111) | 0.290 | 0.343 |
| (222) and (111) | 0.975 | 0.094 |
| (400) and (111) | 0.206 | 0.041 |

**Table.6**. Particle size of Lead nanopowder

| 2θ of the intense peak (deg) | hkl | FWHM of Intense peak (β) radians | Size of the partcle (D) nm | d-spacing Å | Lattice parameter (*a*) Å |
|---|---|---|---|---|---|
| 31.3777 | (111) | 0.002918 | 49 | 2.85096 | 4.938 |
| 36.3611 | (200) | 0.00321 | 45 | 2.47086 | 4.942 |
| 52.3101 | (220) | 0.00233 | 66 | 1.74894 | 4.947 |
| 62.2696 | (311) | 0.00584 | 28 | 1.49102 | 4.945 |
| 65.3477 | (222) | 0.00584 | 28 | 1.42804 | 4.947 |
| 77.0153 | (400) | 0.00712 | 25 | 1.23719 | 4.949 |

A good agreement between the Experimental diffraction angle [2θ] and Standard diffraction angle [2θ] of specimen is confirming standard of the specimen [33]. Six peaks at 2 θ values of 31.3777, 36.3611, 52.3101, 62.2696, 65.3477 and 77.0153 deg corresponding to (111), (200),

(220) (311), (222) and (400) planes of lead is observed and compared with the standard powder diffraction card of Joint Committee on Powder Diffraction Standards (JCPDS), lead file No. 65-2873. The d-spacing values of experimental is also confirming to the standard values. The XRD study confirms / indicates that the resultant particles are (FCC) lead nanopowder and the details are presented in the Table.4

It is worth noting that the ratio between the intensities of the diffraction peaks [34]. The ratio between peaks (200) & (111), (220) & (111), (311) & (111), (222) & (111) and (400) & (111) are enumerated in Table.5. These values also close with the conventional value.

### 3.2. XRD – Lattice Constant

The FCC crystal structure of lead has unit cell edge '$a$' = 4.9497 Å and this value is calculated theoretically by using formula,

$$a = 4/\sqrt{2} \times r \quad \text{............……………………………….....} (4)$$

For lead r = 0.175 nm. The experimental lattice constant '$a$' is calculated from the most intense peak (111) of the XRD pattern is 4.938 Å. Both theoretical & experimental lattice constant '$a$' are in agreement. The details of '$a$' value of all peaks have been produced in Table.6. The unit cell volume calculated from experimental '$a$' is 120.4074Å$^3$.

### 3.3. XRD - Particle Size Calculation

From this study, considering the peak at degrees, average particle size has been estimated by using *Debye-Scherrer formula*.

$$D = \frac{0.9\lambda}{\beta \cos\theta} \quad \text{........……………..……………………......} (5)$$

Where '$\lambda$' is wave length of X-Ray (0.1540 nm), '$\beta$' is FWHM (full width at half maximum), '$\theta$' is the diffraction angle and 'D' is particle diameter size. The calculated particle size details are in Table.6. The value of d (the interplanar spacing between the atoms) is calculated using *Bragg's Law*.

$$2d\sin\theta = n\lambda \quad \text{…………………………………………..……...} (6)$$

### 3.4. XRD - Expected 2θ Positions

The *expected 2θ positions of all the six peaks* in the diffraction pattern and the *interplanar spacing d* for each peak are calculated using following formula and the details are shown in Table.7.

$$1/d^2 = h^2 + k^2 + l^2 / a^2 \quad \text{………………...…….} (7)$$

*Bragg's Law* is used to determine the 2θ value: The expected 2θ and d values are close with the experimental 2θ and d values.

**Table 7.** The lattice plane and the lattice spacing-d from XRD

| hkl | 2θ (deg) | | d (Å) | |
|---|---|---|---|---|
| | Experiment | Expected | Experiment | Expected |
| 111 | 31.3777 | 31.2835 | 2.85096 | 2.8577 |
| 200 | 36.3611 | 36.2790 | 2.47086 | 2.4749 |
| 220 | 52.3101 | 52.2448 | 1.74894 | 1.74998 |
| 311 | 62.2696 | 62.1669 | 1.49102 | 1.4924 |
| 222 | 65.3477 | 65.2642 | 1.42804 | 1.4289 |
| 400 | 77.0153 | 77.0219 | 1.23719 | 1.2374 |

### 3.5. XRD - Instrumental Broadening

When particle size is less than 100 nm, appreciable broadening in x-ray diffraction lines will occur. Diffraction pattern will show broadening because of particle size and strain. The observed line broadening will be used to estimate the average size of the particles. The total broadening of the diffraction peak is due to sample and the instrument. The sample broadening is described by

$$FW(S) \times \cos\theta = \frac{K \times \lambda}{Size} + 4 \times Strain \times \sin\theta \qquad (8)$$

The total broadening $\beta_t$ is given by the equation

$$\beta_t^2 \approx \left\{\frac{0.9\lambda}{D\cos\theta}\right\}^2 + \{4\varepsilon\tan\theta\}^2 + \beta_0^2 \qquad (9)$$

$\varepsilon$ is strain and $\beta_0$ instrumental broadening. The average particle size D and the strain $\varepsilon$ of the experimentally observed broadening of several peaks will be computed simultaneously using *least squares method*. Instrumental Broadening is presented in Figure.3. Due to the size effect, the peaks broaden and then widths become larger as the particle size becomes smaller. The broadening of the peak may also occur due to micro strains of the crystal structure arising from defects like dislocation and twinning [35]

Williamson and Hall proposed a method for deconvoluting size and strain broadening by looking at the peak width as a function of 2θ. Here, Williamson-Hall plot is plotted with sin θ on the x-axis and $\beta$ cos θ on the y-axis (in radians). A linear fit is got for the data. From the linear fit, particle size and strain are extracted from y-intercept and slope respectively. The extracted particle size is 8 nm and strain is 0.0021. Table.8 provides all the data details and Figure.4 shows Williamson Hall Plot.

Line broadening analysis is most accurate when the broadening due to particle size effects is at least twice the contribution due to instrumental broadening. The size range is calculated over which this technique will be most accurate. A rough upper limit is estimated for reasonable accuracy by looking at the particle size that would lead to broadening equal to the instrumental broadening. For example, for Monochromatic Lab X-ray (Cu Kα FWHM ~ 0.05° at 20° 2θ), the accurate Size Range is < 90 nm (900 Å) and the rough Upper Limit is = < 180 nm (1800 Å).

**Table 8.** Data for Insrumental Broadening and W.H. Plot

| 2θ (deg) | θ (deg) | FWHM-β (Radian) | sin θ | cos θ | β cos θ |
|---|---|---|---|---|---|
| 31.3777 | 15.6888 | 0.002918 | 0.2704 | 0.9627 | 0.00281 |
| 36.3611 | 18.1805 | 0.00321 | 0.3120 | 0.9500 | 0.00305 |
| 52.3101 | 26.1550 | 0.00233 | 0.4408 | 0.8976 | 0.00209 |
| 62.2696 | 31.1348 | 0.00584 | 0.5171 | 0.8559 | 0.004999 |
| 65.3477 | 32.6738 | 0.00584 | 0.5399 | 0.8417 | 0.00492 |
| 77.0153 | 38.5076 | 0.00712 | 0.6226 | 0.7825 | 0.00557 |
| y intercept = 1E-4 ±0.0041 | | | size = 8 nm | | |
| slope = 0.0084 ±0.0087 | | | strain = 0.0021 | | |

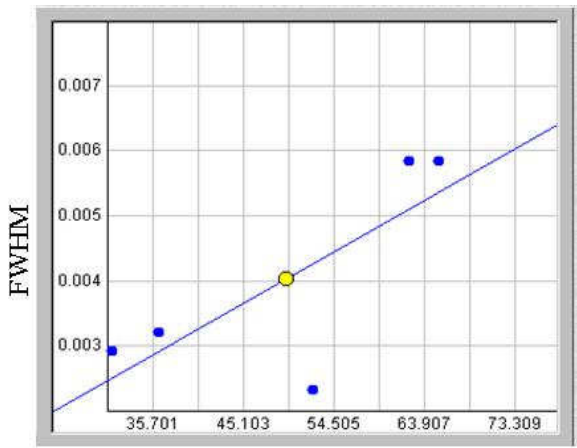
**Fig.3.** Typical Instrumental Broadening.
y = 0.000084x - 0.000155

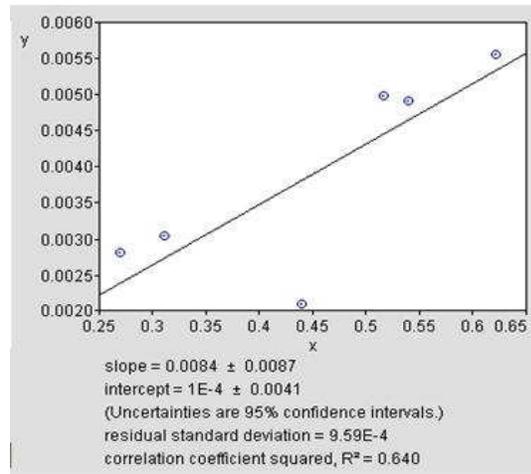
**Fig.4**. Williamson Hall Plot is indicating line broadening value due to the equipment.

### 3.6. XRD – Dislocation Density

The dislocation density is defined as the length of dislocation lines per unit volume of the crystal. [36]. In materials science, a dislocation is a crystallographic defect, or irregularity, within a crystal structure. The presence of dislocations strongly influences many of the properties of materials. Mathematically, dislocations are a type of topological defect. The dislocation density increases with plastic deformation, a mechanism for the creation of dislocations must be activated in the material. Three mechanisms for dislocation formation are formed by homogeneous nucleation, grain boundary initiation, and interface the lattice and the surface, precipitates, dispersed phases, or reinforcing fibers.

The movement of a dislocation is impeded by other dislocations present in the sample. Thus, a larger *dislocation density* implies a larger hardness. Chen and Hendrickson measured and determined dislocation density and hardness of several silver crystals. They found that crystals with larger dislocation density were harder [37]. It has been shown for different pure face-centered cubic (fcc) metals processed by Equal Channel Angular Pressing (ECAP) that the dislocation density increases while the grain size decreases with increasing strain and ultimately these parameters reach saturation values [38].

It is well known that above a certain grain size limit (~20 nm) the strength of materials increases with decreasing grain size [39, 40]. The X-ray line profile analysis has been used to determine the intrinsic stress and dislocation density [41, 42]. The dislocation density (δ) in the sample has been determined using expression [43].

$$\delta = \frac{15\beta \cos\theta}{4aD} \quad \ldots (10)$$

Where δ is dislocation density, $\beta$ is broadening of diffraction line measured at half of its maximum intensity (in radian), $\theta$ is Bragg's diffraction angle (in degree), $a$ is lattice constant (in nm) and $D$ is particle size (in nm). The dislocation density can also be calculated from

$$\delta = 1/D^2 \quad \ldots (11)$$

Where δ is dislocation density and D is the crystallite size [44]. Results of the dislocation density calculated from both the formulas are approximately same. The number of unit cell is calculated from

$$n = \pi(4/3) \times (D/2)^3 \times (1/V) \quad \ldots (12)$$

Where D is the crystallite size and V is the cell volume of the sample [45].

**Table.9**. Dislocation Density and Number of Unit Cell from XRD

| 2θ (deg) | Particle Size D (nm) | Dislocation Density (m$^2$) | | Number of Unit Cell |
|---|---|---|---|---|
| | | δ = 15βcosθ / 4aD | δ = 1 / D$^2$ | |
| 31.3777 | 49 | 4.35 x 10$^{14}$ | 4.16 x 10$^{14}$ | 5.11 x 10$^5$ |
| 36.3611 | 45 | 5.14 x 10$^{14}$ | 4.94 x 10$^{14}$ | 3.95 x 10$^5$ |
| 52.3101 | 66 | 2.41 x 10$^{14}$ | 2.30 x 10$^{14}$ | 12.43x 10$^5$ |
| 62.2696 | 28 | 1.35 x 10$^{15}$ | 1.28 x 10$^{15}$ | 0.95 x 10$^5$ |
| 65.3477 | 28 | 1.33 x 10$^{15}$ | 1.28 x 10$^{15}$ | 0.95 x 10$^5$ |
| 77.0153 | 25 | 1.68 x 10$^{15}$ | 1.60 x 10$^{15}$ | 0.68 x 10$^5$ |

The dislocation density of sample lead nanoparticles is enumerated in Table.9. It is observed from these tabulated details and from Figures.5, 6 & 7, dislocation density is indirectly proportional to particle size and number of unit cell. Dislocation density increases while both particle size and number of unit cell decreases.

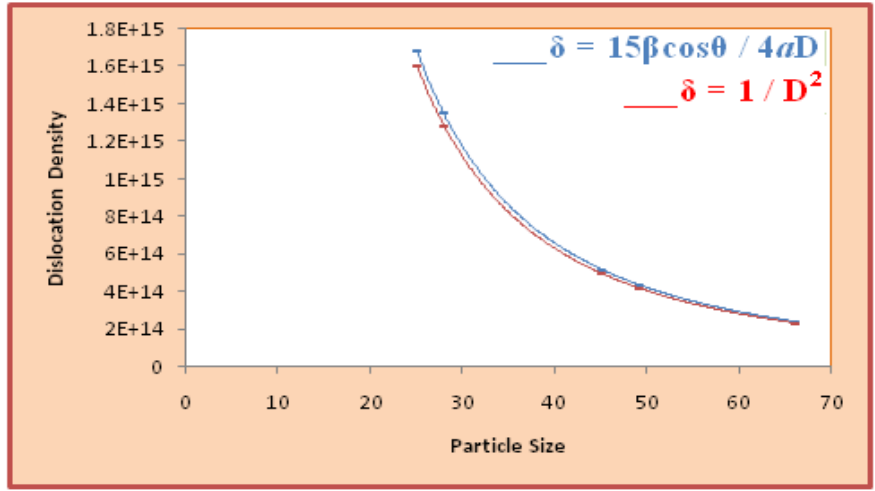

**Fig.5.** Dislocation Density Vs Particle Size

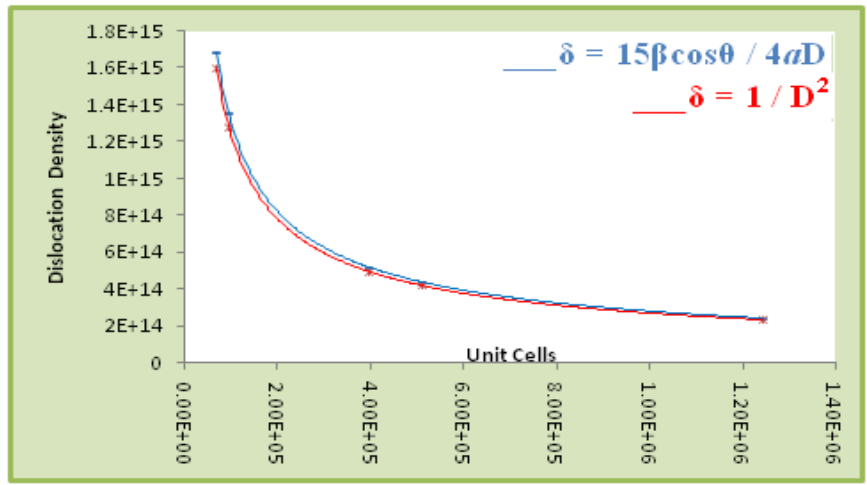

**Fig.6.** Dislocation Density Vs Number of Unit Cells

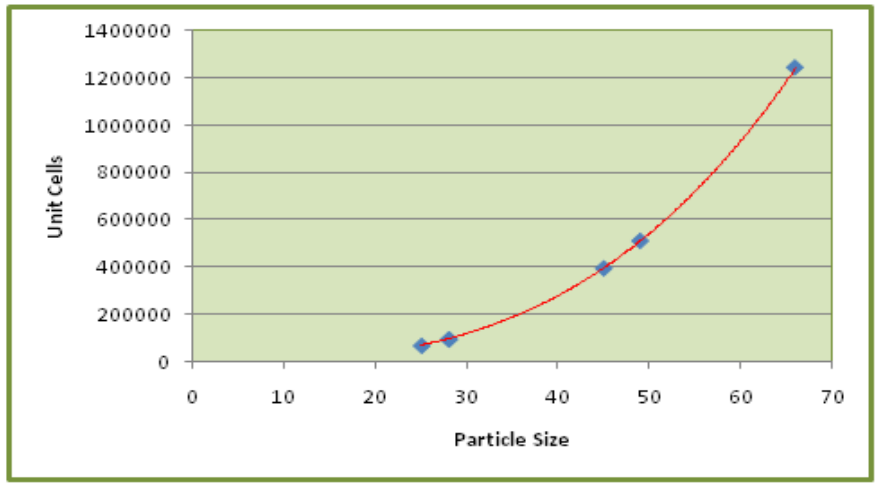

**Fig.7.** Number of Unit Cells Vs Particle Size

On the other side, particle size increases while number of unit cell increases. The ratio between increasing and decreasing value is not in linear. The average dislocation density of lead nanoparticles found to be as $11.1 \times 10^{14}$ m$^{-2}$. The dislocation density of electrolyzed silver nanoparticles $9.2 \times 10^{14}$ m$^{-2}$ has been noted from our earlier report [46]. While comparing both these values, it is observed that the sample lead nanoparticles harder than electrolyzed silver nanoparticles.

### 3.7. XRD – Theoretical Density

The theoretical density was calculated using the lattice parameters with formula [47, 48]

$$D_{th} = Z \frac{M}{N \times V} \quad \quad \quad (13)$$

where 'M' (in atomic-weight units) is 207.8 - the mass of one unit of the chemical formula, 'z' is 4 the number of such chemical units in one unit cell of the crystal, 'N' is the Avagadro's number and V is 120.4074Å$^3$ - the volume of the crystalline unit cell as determined by XRD. The calculated theoretical density value is 11.461 g/cm$^3$ and it is agreed well with the reported data.

### 3.8. XRD - Crystallinity Index

It is generally agreed that the peak breadth of a specific phase of material is directly proportional to the mean crystallite size of that material. Quantitatively speaking, sharper XRD peaks are typically indicative of high nanocrystalline nature and larger crystallite materials. From our XRD data, a peak broadening of the nanoparticles is noticed. The average particle size, as determined using the Scherrer equation, is calculated to be 8 nm. Crystallinity is evaluated through comparison of crystallite size as ascertained by TEM particle size determination. crystallinity index Eq. is presented below:

$$I_{cry} = \frac{D_p(SEM, TEM)}{D_{cry}(XRD)} (I_{cry} \geq 1.00) \quad \quad \quad (14)$$

Where $I_{cry}$ is the crystallinity index; $D_p$ is the particle size (obtained from either TEM or SEM morphological analysis); $D_{cry}$ is the particle size (calculated from the Scherrer equation). If $I_{cry}$ value is close to 1, then it is assumed that the crystallite size represents monocrystalline whereas a polycrystalline have a much larger crystallinity index [49]. The crystallinity index of the sample is 1.25 which is more than 1.0. The details are enumerated in Table.10 which indicates that the sample lead nanoparticles are highly crystalline and fcc phase structure is well-indexed.

Table.10. The crystallinity index of Lead Nanoparticles

| Sample | Dp (nm) | Dcry (nm) | Icry (unitless) | Particle Type |
|---|---|---|---|---|
| Lead Nanoparticles | 10 | 8 | ~1.25 | Polycrystalline |

## 3.9. XRD – Morphology Index

It is well known that lead powder is widely used in many diverse industries such as scientific, nuclear Oil & gas exploration, anti-corrosive paints, lubricating, and sinter materials, catalysts, explosives industries and also in medical, electrical fields. The use of lead nanopowder is derived from its unique structural, physical and chemical properties, which are reflected by its hardness, surface properties, particle size and morphology. It is proposed that the specific surface area of lead nanopowder (which is important to many of the above mentioned industries) is dependent on the interrelationship of particle morphology and size. A XRD morphology index (MI) is calculated from FWHM of XRD data to understand this relationship, based on our earlier report [50]. MI is obtained using equation.

$$M.I. = \frac{FWHM_h}{(FWHM_h + FWHM_p)} \quad \ldots\ldots\ldots\ldots\ldots\ldots (15)$$

Where M.I. is morphology index, FWHMh is highest FWHM value obtained from peaks and FWHMp is value of particular peak's FWHM for which M.I. is to be calculated.

Table.11. Morphology Index of Nano-sized Lead particles

| FWHM (β) radians | Particle Size (D) nm | Surface Area (nm$^2$) | Volume (nm$^3$) | Specific Surface Area (m$^2$ g$^{-1}$) | Morphology Index (unitless) |
|---|---|---|---|---|---|
| 0.002918 | 49 | 99982 | 1178588 | 7.401 | 0.7093 |
| 0.00321 | 45 | 91538 | 994021 | 8.072 | 0.6892 |
| 0.00233 | 66 | 136433 | 2138251 | 5.567 | 0.7530 |
| 0.00584 | 28 | 56209 | 384846 | 12.743 | 0.5493 |
| 0.00584 | 28 | 56209 | 384846 | 12.743 | 0.5493 |
| 0.00712 | 25 | 50069 | 306796 | 14.239 | 0.5000 |

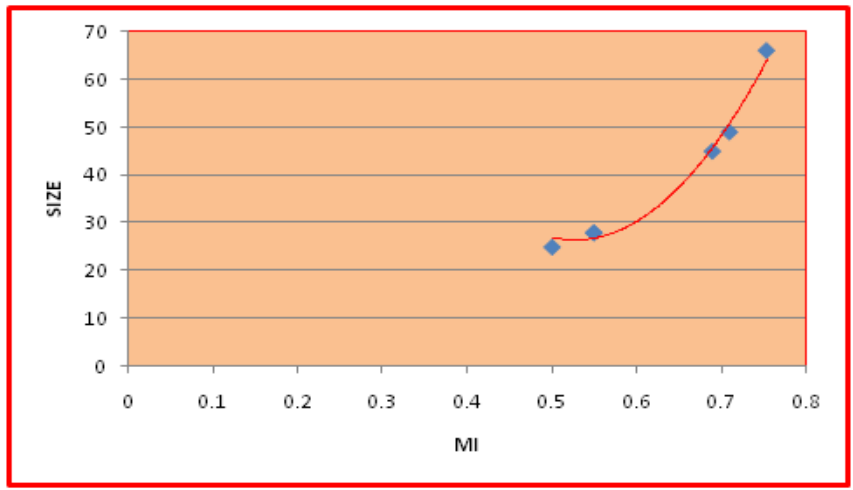

**Fig.8.** Morphological Index Vs Particle Size

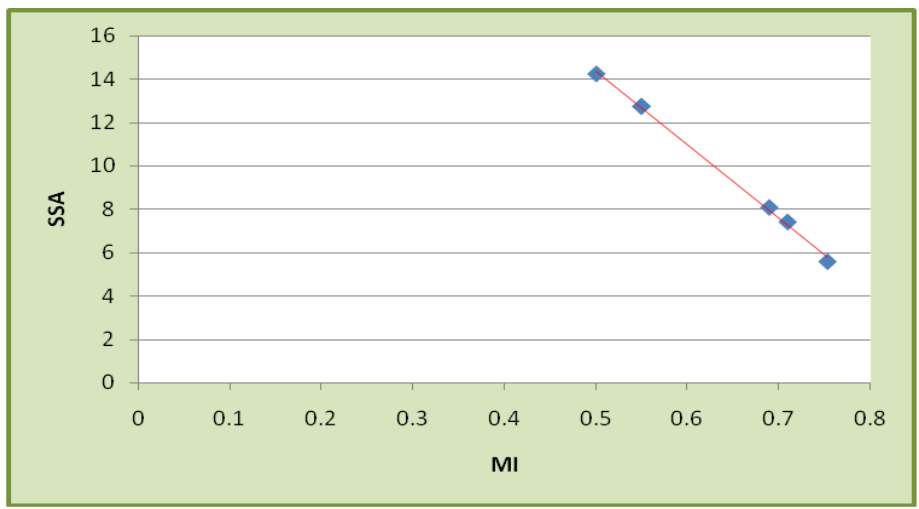

**Fig.9.** Morphological Index Vs Specific Surface Area

Experimental lead nanopowder MI range is from 0.50 to 0.75 and the details are presented in Table.11. It is correlated with the particle size (range from 25 to 66 nm) and specific surface area (range from 5.567 to 14.239 m$^2$g$^{-1}$). It is observed that MI has direct relationship with particle size and an inverse relationship with specific surface area. The results are shown in Fig.8 & 9. A linear fit in MI Vs SSA diagram indicates the good co-ordination between them.

### 3.10. XRD – Unit Cell Parameters

Unit cell parameters values calculated from XRD are enumerated in table.12.

**Table.12**. XRD parameters of Lead Nanoparticles

| Parameters | Values |
|---|---|
| Structure | FCC |
| Space group | Fm-3m (Space group number: 225) |
| Point group | m3m |
| Packing fraction | 0.74 |
| Symmetry of lattice | cubic close-packed |
| Particle size | 8 nm |
| Bond Angle | α = β = γ = 90° |
| Lattice parameters | a = b = c = 4.938 Å |
| Vol.unit cell(V) | 120.4074 Å$^3$ |
| Radius of Atom | 175 pm |
| Density ( ρ ) | 11.461 g cm$^{-3}$ |
| Dislocation Density | 11.1 x 10$^{14}$ m$^{-2}$ |
| Mass | 207.2 amu |

### 3.11. SEM Analyses Pb Nanorods

Metal nanowires are one of the most attractive materials. They are used as interconnects for nanoelectronics, magnetic devices, chemical / biological sensors, and biological labels. Wires with large aspect ratios >20 (length-to diameter ratio) are called nanowires and those with small aspect ratios are called nanorods. "Wires" are bridged between two larger electrodes; they are often referred to as nanocontacts [51]. The SEM images have shown the result in Figure.10 (a-d). The images reveal about the changed shape i.e. rod shaped Pb nanoparticles.The length to diameter ratio from SEM analysis; largest rod is 10 and smallest rod is 5. These ratio < 20 indicates rod shape.

Size and shape control many of the properties of nanomaterials like physical, chemical, catalytic, optical etc. Solvent, pH value, photo-sensitizer, the concentration and category of stabilizer, irradiation intensity, wavelength, and irradiation time are all important factors for shape and size controlled synthesis of metal nanoparticles.

From earlier reports; gold nanorods have been intensively affected by the wavelength of UV light source and their photon flux; rod like gold particles have been prepared by UV irradiation; Metals have strongly absorb light in the visible region due to surface plasma resonance; the number of surface amino group of dendrimers have a key factor to control the particles with UV irradiation method; While on synthesis of Pb nanoparticles, increasing the concentration of surfactant / stabilizing agent have produced Pb hollow nanoparticles and decreasing its concentration have produced solid Pb nanoparticles.

On this basis of these reports, an attempt has been made to change the shape of Pb nanoparticles (konjac extract added) from spherical to rod by the influence of sunbeams. The result explicates that it is possible to change the shape even after completion of the entire synthesis process. The SEM Fig.10 (d) shows some spherical particles among the lead nanorods which indicates that these rods have been made up of spherical particles (nucleation growth of particles). The rods have been formed by the binding of spherical particles one by one (one over one) and have been elongated in one direction after binding. Due to this, some up and down structures are in the outer surfaces of the nanorods. The surfaces (textures) are not uniform and not smooth.

Sunlight is the total frequency spectrum of electromagnetic radiation given off by the Sun, particularly infrared, visible, and ultraviolet light. Bright sunlight provides illuminance of approximately 100,000 lux or lumens per square meter. Pb nanoparticles strongly absorb light in the visible region due to Surface Plasmon Resonance (SPR). The absorbed light promotes the reactions / effects of the sunlight on Pb nanoparticles, in presence of the amino group of konjac aqueous extract which elongates the nanoparticles in one direction and change their shape from spherical to rod.

The synthesized nanorods are in various sizes. The length is varying from 0.5 $\mu$m to 12 $\mu$m and diameter is 0.1 $\mu$m to1.25 $\mu$m. It has been considered to calculate, the average size between smallest rod and largest rod, to ascertain the size of nanorod. The length of largest rod is 12 $\mu$m and diameter is 1.25 $\mu$m. The length of smallest rod is to 0.5 $\mu$m and diameter is to 0.1 $\mu$m. The average length of rod is to 6.25 $\mu$m and diameter is to 0.675 $\mu$m. However, SEM analysis does not give real image and it gives real shape only. The above estimated average size of rod is not exact one and it is possible from TEM analysis only.

It is apparently that the diameter of the smallest rod is equal to the diameter of spherical particle because the rods have been grown by nucleation growth of smaller particles. It is compared, the diameter of the smallest rod (0.1 $\mu$m) from SEM image with particle diameter (10 nm) from TEM image. From this comparison, it is found that the rod diameter 10 times more than TEM particle diameter and the average of the rod has been divided by 10 to assess the real average size of the rod. It is found that the real average length of rod is 625 nm and diameter is 67 nm. This observed value 67 nm corroborates with the particle size-diameter 66 nm from (220) indexed peak of XRD. Also, it is close to the rod diameter 55 nm calculated by particle size analyzer.

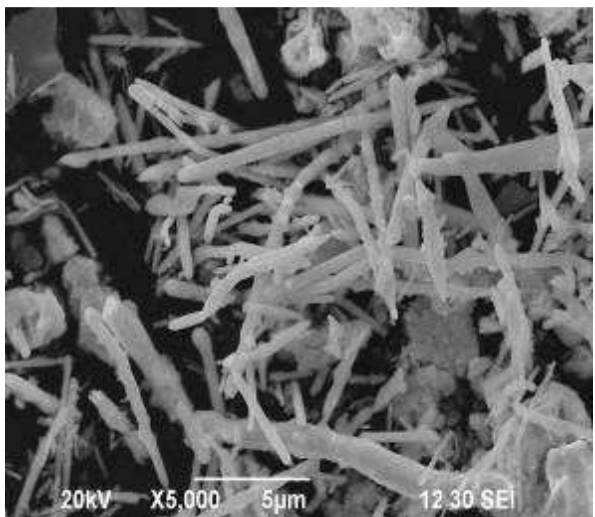
**Fig.10(a).** Pb Nanorods 5000 magnification.

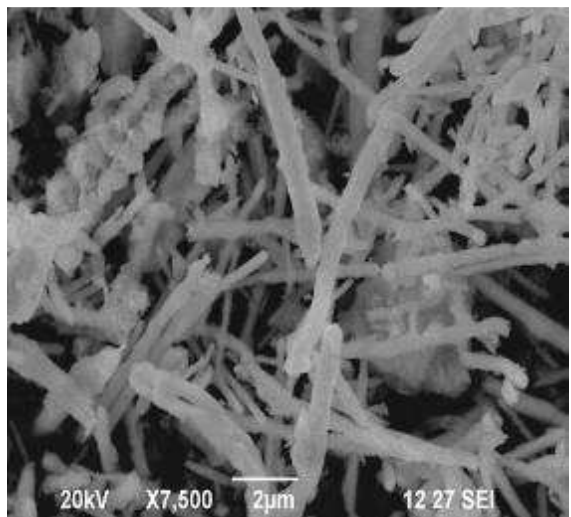
**Fig.10(b).** Pb Nanorods 7500 magnification

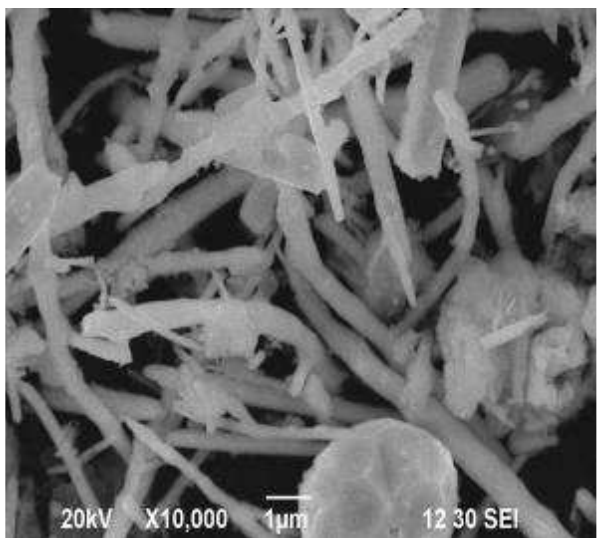
**Fig.10(c).** Pb Nanorods 10000 magnification

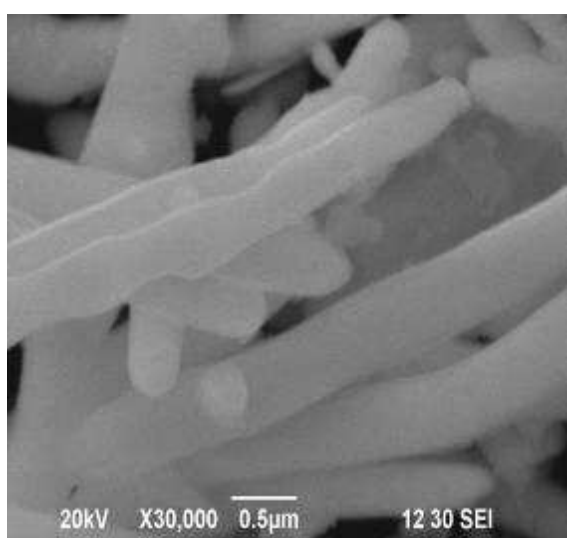
**Fig.10(d)**. Pb Nanorods 30000 magnification

### 3.12. Specific Surface Area

The surface states will play an important role in the nanoparticles, due to their large surface to volume ratio with a decrease in particle size [52]. Specific surface area (SSA) is a material property. It is a derived scientific value that can be used to determine the type and properties of a material. It has a particular importance in case of adsorption, heterogeneous catalysis and reactions on surfaces. SSA is the SA per mass.

$$SSA = \frac{SA_{part}}{V_{part} * density} \quad \ldots\ldots\ldots\ldots\ldots\ldots\ldots\ldots\ldots (16)$$

Here Vpart is particle volume and SApart is Surface Area of particle [53].

$$S = 6 * 10^3 / D_p \rho \quad \ldots\ldots\ldots\ldots\ldots\ldots\ldots\ldots\ldots (17)$$

Where S is the specific surface area, Dp is the size of the particles (spherical shaped), and $\rho$ is the density of Pb 11.461 gcm$^{-3}$ [54]. Mathematically, SSA can be calculated using these formulas. Both of these formulas yield same result. For SSA of rod shaped particles, 4 should be used in place of 6 in eq.17 ( $S = 4 * 10^3 / D_p \rho$ ). The nanorod length and diameter estimated by SEM analysis has been used to calculate SSA and details are in Table.13.

**Table.13.** Specific Surface Area of Lead Nanoparticles

| Particle Size (nm) | | Surface Area (nm$^2$) | Volume (nm$^3$) | Density (g cm$^{-3}$) | SSA (m$^2$g$^{-1}$) | SA to Volume Ratio |
|---|---|---|---|---|---|---|
| Length | Diameter | | | | | |
| 625 | 67 | 139693 | 2236540 | 11.461 | 5.4497 | 0.0624 |

## 3.13. TEM Analyses of spherical Pb Nanoparticles

HRTEM pictures of the synthesized lead nanoparticles have been shown in Figure.11 (a-e). Figure.11 (b) shows the diameter of the particle is approximately 10 nm. Figure 11(f) is STEM mode image taken with post filter HAADF detector (High Angle Annular Dark Field). It shows the spot size is 8nm. (1 0 0) face of FCC structure model created by BALSAC surface explorer has been shown in Figure.11 (g). Selected Area Electron Diffraction (SAED) pattern has been taken over 200 micron (diameter) condenser apertures is in Figure.11 (h). Indexed SAED pattern is in Figure.11 (i) which confirms FCC structure of the lead. SAED ring pattern is similar to XRD Debye-Scherrer pattern. It provides information about the crystal structure and lattice spacing of the sample; however this information will not be as accurate as from XRD. For indexing purpose of SAED pattern, diameters ($2r_n$) of rings were measured, ratios between them ($2r_n$ / $2r1$) were calculated and the details were provided in Table. 14. The diffraction planes (hkl) ratios $M_n$ / $M1$ of $\sqrt{h^2+k^2+l^2}$ match the ratios of diameters ($2r_n$ / $2r1$) which confirms the ring pattern is FCC pattern. Figure.11 (j) shows the SAED part of the Bright Field image (direct beam image) and Dark Field image (diffracted beam image). HRTEM image of a particle is shown in Figure.11 (n). It shows the surfaces and FCC structure of the synthesized lead nanoparticle. Figure.11 (o) shows the d-spacing 2.83 Å which is well agreement with XRD d-spacing 2.85 Å of (1 1 1) indexed, most intense peak at 2θ 31.37. A zoomed portion of the lattice fringes (whitish grey colour lines) of that particle is shown in Figure.11 (p).

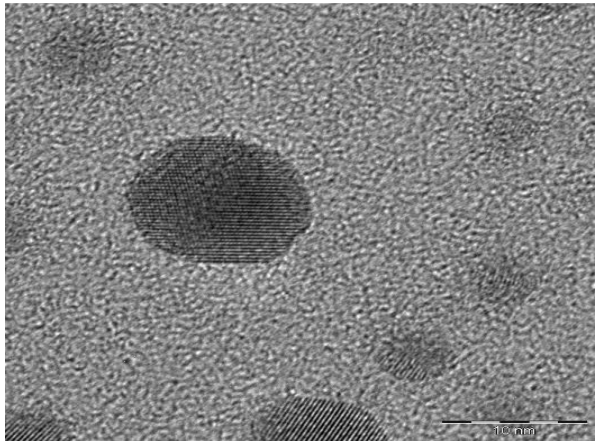
**Fig.11 (a).** HRTEM - Nanospherical Pb

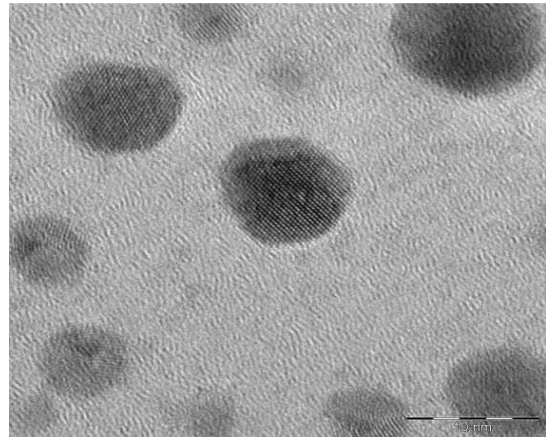
**Fig.11 (b).** HRTEM - Nanospherical Pb

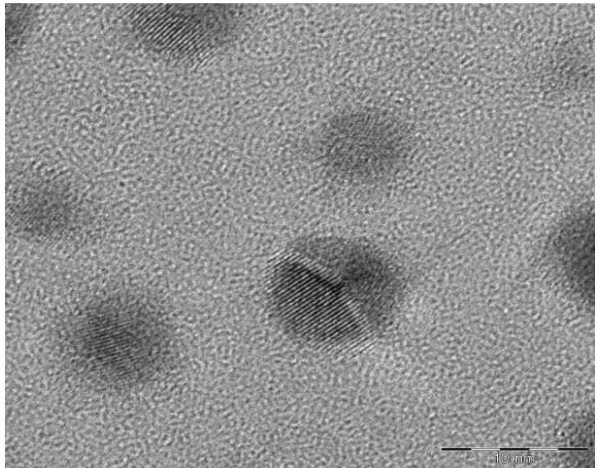
**Fig.11 (c).** HRTEM - Nanospherical Pb

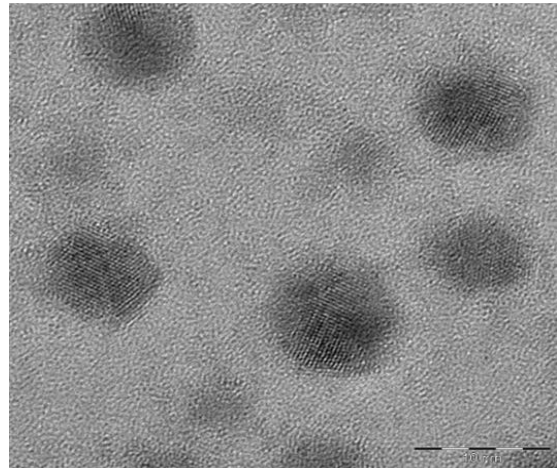
**Fig.11 (d).** HRTEM - Nanospherical Pb

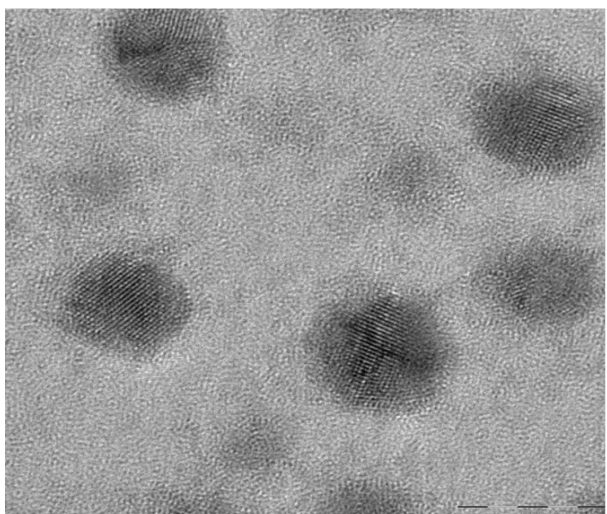
**Fig.11 (e).** HRTEM - Nanospherical Pb

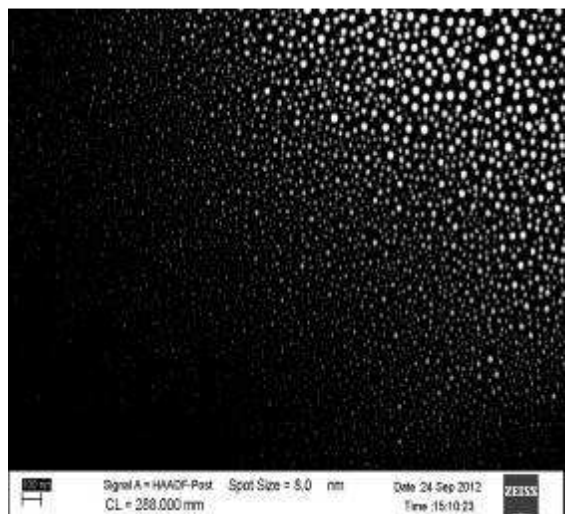
**Fig.11 (f).** HAADF - Nanospherical Pb

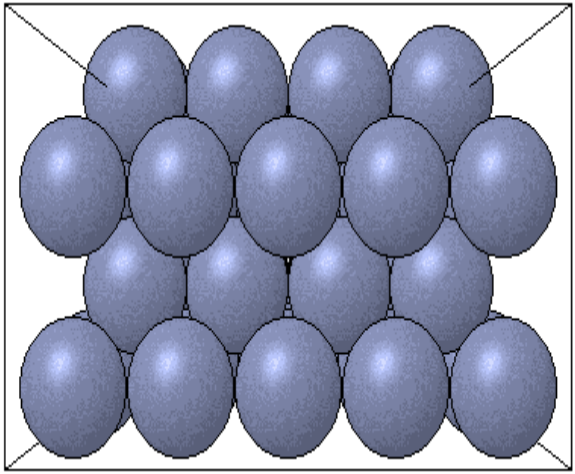
**Fig.11 (g).** FCC (100) face by BALSAC

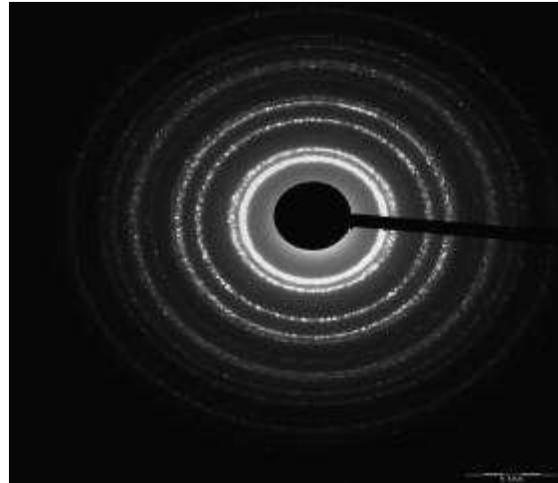
**Fig.11 (h).** SAED Image Nanospherical Pb

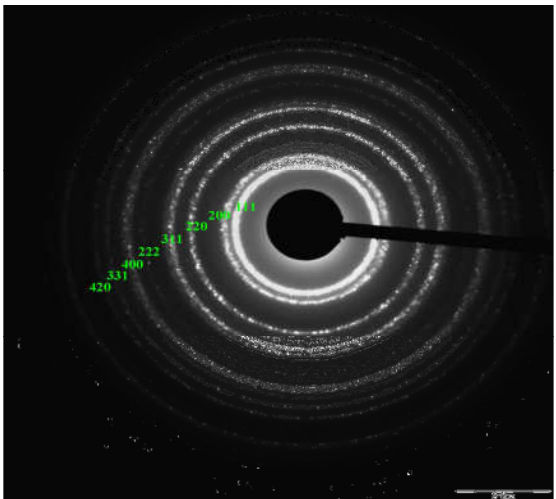
**Fig.11 (i).** Indexed – SAED Image

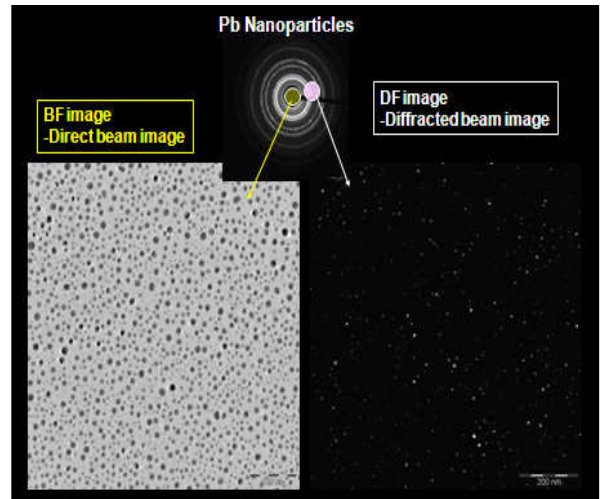
**Fig.11 (j).** BF & DF Images of SAED Pattern

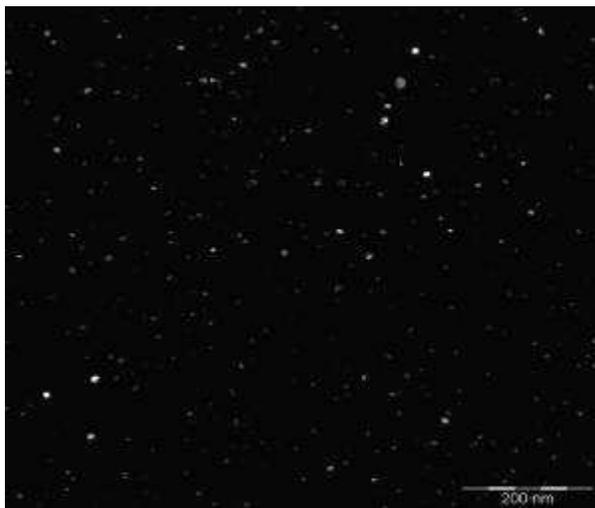
**Fig.11 (k).** Pb nanoparticles in Dark Field

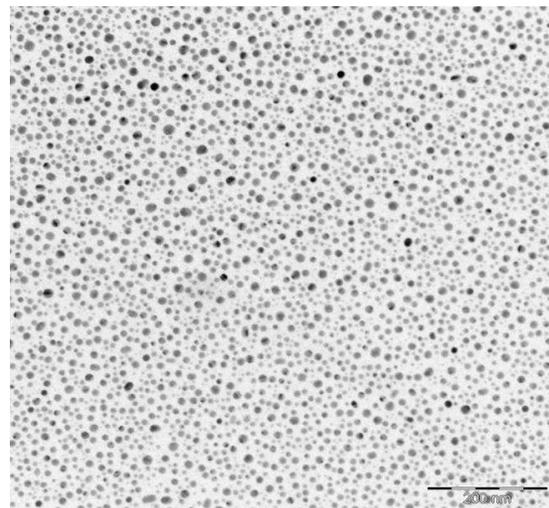
**Fig.11 (l).** Pb nanoparticles in Bright Field

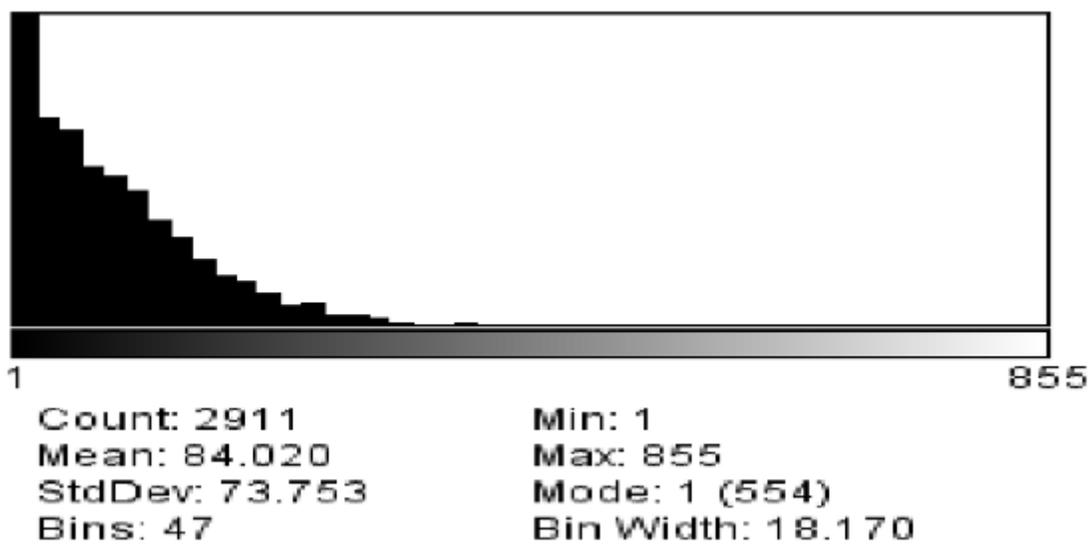

**Fig.11 (m).** Area Distribution Histogram

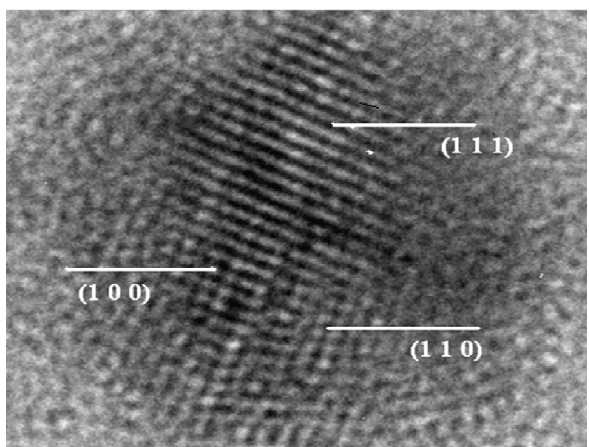

**Fig.11 (n).** Surfaces of Pb Nanoparticle

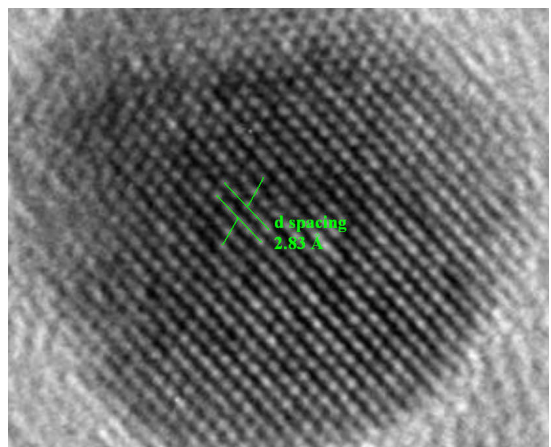

**Fig.11 (o).** d-spacing of Nanospherical Pb

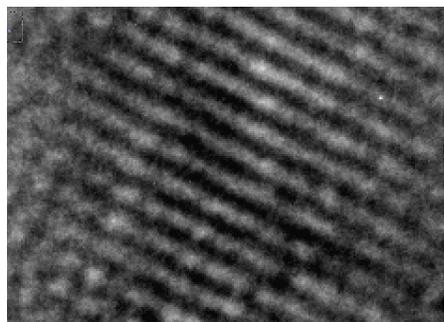

**Fig.11 (p).** Zoomed Lattice Fringes

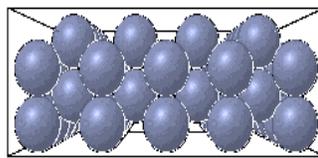

**Fig.11 (q).** FCC (110) face

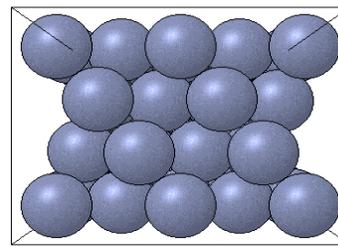

**Fig.11 (r).** FCC (111) face

Surface models of (1 1 0) and (1 1 1) of FCC structure are presented in Figure.11 (q) and 11 (r) respectively. Figure.11 (k) and 11 (l) are showing particles distribution in dark field and bright field respectively. From these pictures, it is concluded that the particles are in spherical in nature, uniform shaped and without agglomeration. Particles area distribution histogram is given in the Figure.11 (m). Average area (assuming the particles are round in shape - for calculation purpose only) of nanoparticles is 84.020 $nm^2$. From this average area (A), the average particle diameter (D) 10.34 nm is calculated from using equation

$$D = 2\sqrt{A/\pi} \qquad (18)$$

The average area (A) of spherical shaped nanoparticles 336 $nm^2$ is calculated from the equation

$$A = 4\pi r^2 \qquad (19)$$

Where 'r' is radius of the particle and it is half of 'D'.

**Table.14.** SAED Pattern Rings Measurements

| Allowed hkl | $M = \sqrt{h^2+k^2+l^2}$ | Ratios $M_n/M1$ | Ring Diameter (nm) ($2r_n$) | Ring Radius (nm) ($r_n$) | Ratios of Diameter ($2r_n/2r_1$) |
|---|---|---|---|---|---|
| (111) | 1.73205 | 1 | 77.974 | 38.987 | 1 |
| (200) | 2 | 1.1547 | 90.864 | 45.432 | 1.16 |
| (220) | 2.8284 | 1.6329 | 130.724 | 65.362 | 1.67 |
| (311) | 3.3166 | 1.9148 | 157.108 | 78.554 | 2.01 |
| (222) | 3.46 | 1.9976 | 184.689 | 92.344 | 2.36 |
| (400) | 4 | 2.3094 | 202.272 | 101.136 | 2.59 |
| (331) | 4.353 | 2.5132 | 223.959 | 111.979 | 2.87 |
| (420) | 4.4731 | 2.5825 | 239.794 | 119.897 | 3.07 |

## 3.14. EDS Analyses of Pb Nanoparticles

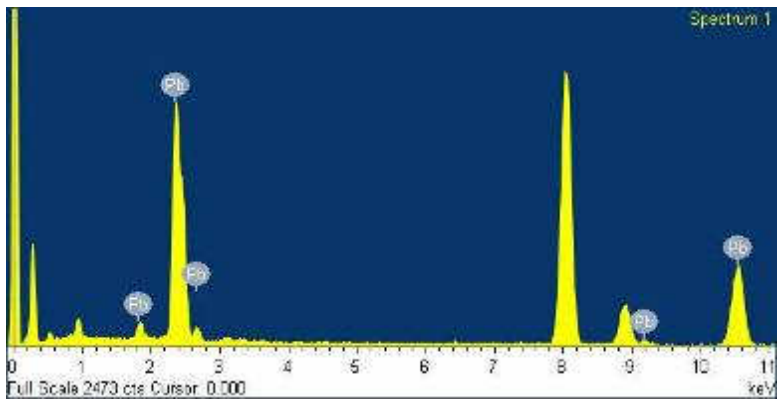
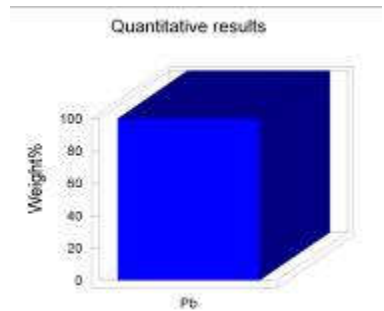

**Fig.12.** EDAX Spectrum showing Pb Nanoparticles    **Fig.13.** Elements % Analyses

The energy value of each peak is matched with X-ray emission wavelength for non-diffractive analysis and the details are presented in Table 15. The EDS analysis results spectrum of lead is shown in Figure.12. It gives distinct elemental signals of lead. The quantitative analysis of lead in weight % abundance is shown figure.13. EDAX analysis is also in well agreement with XRD report and TEM report which indicating that 100% of lead.

**Table.15.** EDS Analysis Result

| Element | Peak Area | Area Sigma | k Factor | Net Counts | Weight % | Atomic % |
|---------|-----------|------------|----------|------------|----------|----------|
| Pb L    |           |            |          |            | 100.00   | 100.00   |
|         | 22915     | 309        | 1.527    | 2473       |          |          |
| Totals  |           |            |          |            | 100.00   |          |

### 3.15. Particle Size Analyses of Pb Nanorods

The particle size analysis report is shown in Fig.14. The rod shape of Pb nanoparticles is further corroborated by the results obtained from DLS experiments. Generally, rod shaped particles will not go under Brownian motion or their motion will too slow to measure and this fact was observed from this analysis. As the Pb nanoparticles were in rod shape, there was no Brownian motion and the size distribution by intensity peak was unable to be measured / seen using this DLS analysis. However, this analysis gives average particle size as 55 nm (Z-Average - d.nm is 1.020e4) and Poly Dispersity Index (PDI) as 0.966. The observed rod size (diameter 55 nm) is close to the rod size (diameter 49 nm) most intensity (111) peak of XRD. A close or equal to 1 of PDI indicates the presence of biological polymer / only one length of polymer. Konjac aqueous extract (bio-polymer) was added to the synthesized lead nanoparticles as stabilizer. PDI 0.966 is observed from this analysis, which indicates the bio-polymer (konjac aqueous extract).

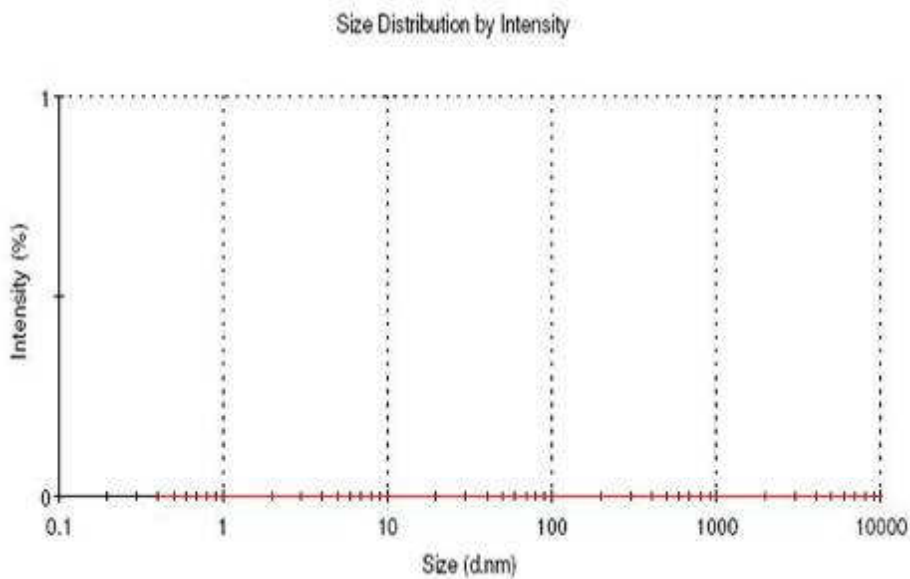

**Fig.14.** Particle Size Analysis

## 3.16. DSC Analyses of Pb Nanorods

The DSC analysis of the lead nanoparticles sample is shown in Fig.15. The heat of fusion and melting point can be determined from the melting curve with pure substances, where the ion temperature side of the melting peak is almost a straight line and the melting point corresponds to the onset. Impure samples often show several peaks. Substances with eutectic impurities exhibit two peaks; first the eutectic peak whose size is proportional to the amount of impurity and then the main melting point. An endothermic peak in a DSC heating curve is a melting peak, if its surface area is between $10 Jg^{-1}$ and $400 Jg^{-1}$.

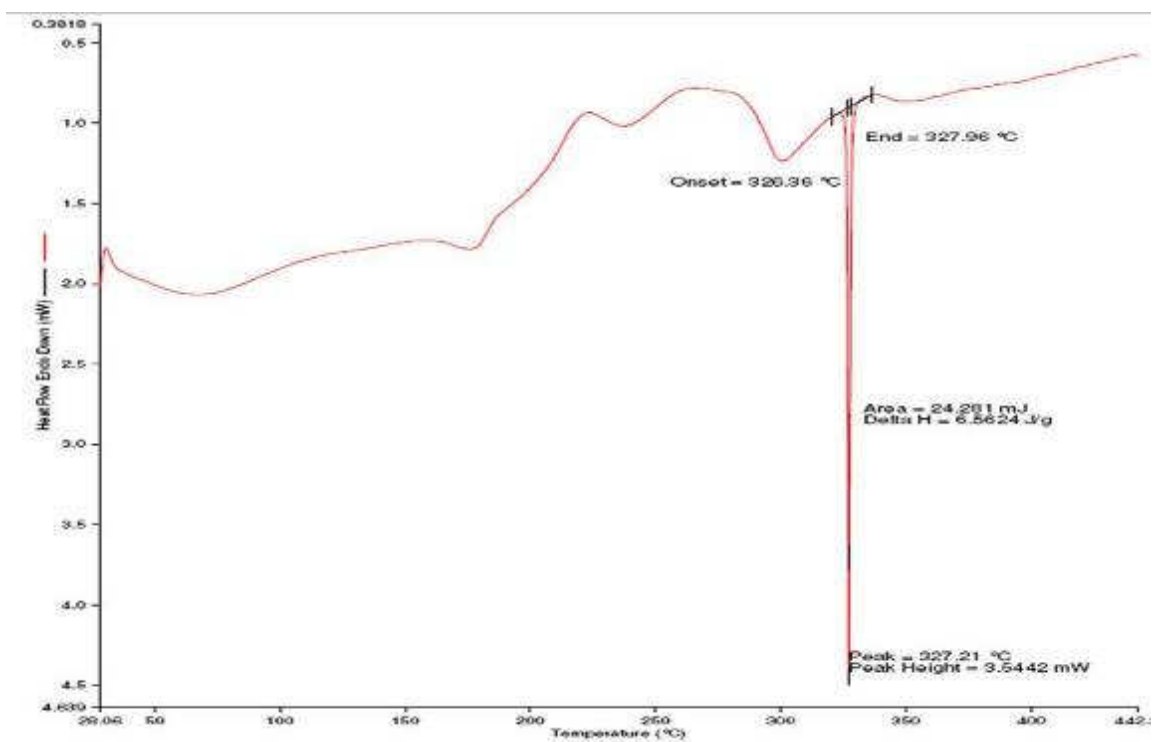

**Fig.15.** DSC Analysis shows Sharp Peak

The onset at 326.36 °C in the DSC analysis report of the sample is the melting point which is well agreement with the lead nanoparticles synthesized by other method. The downward movement of the peak in DSC heating curve indicates that the peak is endothermic peak. The Surface Area (ΔH) has been calculated as 6.5624 $Jg^{-1}$ and this positive value indicates that the observed peak is endothermic peak. The ΔH value is nearer to the beginning of the above said surface area range which forms a melting peak at 327.21 °C for the sample. The melting peak in 327.21 °C appears almost as a straight line. Also, the melting peak is increasingly sharper. These observations show the purity as well as the smaller size of the synthesized sample.

The DSC analysis shows some more peaks which are due to the small layer of Konjac extract over the sample. However, very small peak heights and broaden size (width) of these peaks are indicating that Konjac extract covers only a small layer over the sample.

## 3.17. AAS Analyses of Pb Nanorods

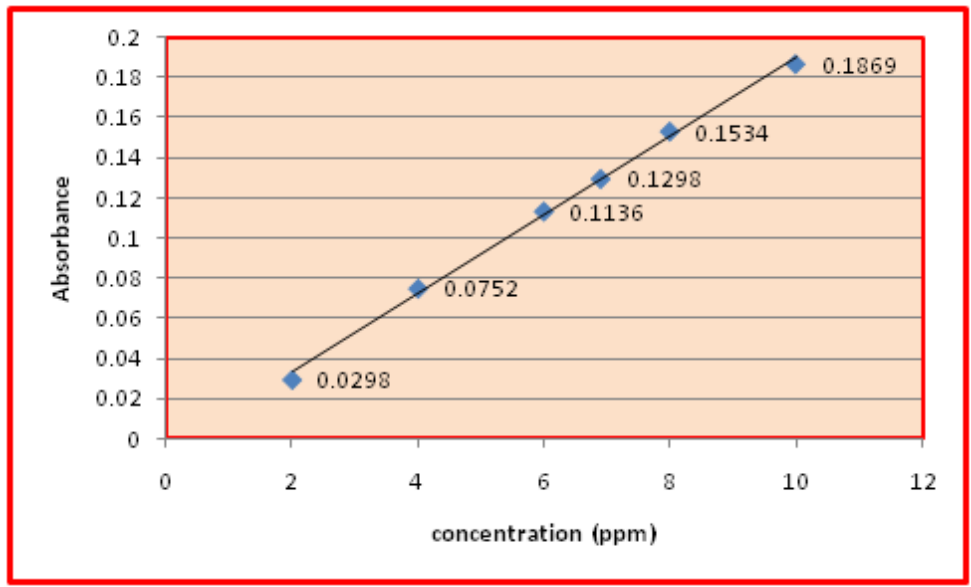

**Fig.16.** Concentration Vs Absorbance

The synthesized lead nanoparticles have been analysed by AAS with optics parameter settings Pb wavelength 283.3 nm and air-acetylene ($C_2H_2$) flame type. The analysis report details are enumerated in Table 16. A calibration curve diagram for Concentration of lead nanoparticles in Parts per Million (ppm) Vs Absorbance has been drawn and a linear fit has been get. It is observed from the Figure.16 that the absorbance is directly proportional to the concentration. The linear fit indicates that the lead nanoparticles have been distributed in proper proportion. Concentration of the sample is 6.9006 ppm (6.9006 mg/l) and absorbance is 0.1298. Absorbance to concentration ratio of the standard at 4 ppm concentration is 0.01880 and for sample is 0.01880. Both the value is same. Likewise, concentration to absorbance ratio of the standard at 4 ppm concentration is 53.1635 and for sample are 53.1633. It indicates high purity of the sample.

**Table16.** AAS Analysis Result

| Standard | | | | | Sample | | | |
|---|---|---|---|---|---|---|---|---|
| True Value | Concentration | Absorbance | BG | %R | Actual Concentration % | Concentration-ppm | Aborbance | BG |
| 2.0000 | 1.5843 | 0.0298 | -0.0010 | 79 | | | | |
| 4.0000 | 3.9979 | 0.0752 | -0.0013 | 100 | 84.5660 | 6.9006 | 0.1298 | -0.0012 |
| 6.0000 | 6.0393 | 0.1136 | -0.0010 | 100.7 | Weight Factor = 0.081600 | | | |
| 8.0000 | 8.1552 | 0.1534 | -0.0008 | 102 | Volume Factor = 100.00 | | | |
| 10.000 | 9.9362 | 0.1869 | -0.0013 | 99.4 | Dilution Factor = 100.00 | | | |
| | | | | | Correction Factor = 0.000100 | | | |

### 3.18. FT-IR Analyses of Pb Nanorods

To demonstrate the Konjac extract bound to the surface of the lead nanoparticle, FTIR analyses were performed. The FT-IR spectrum of lead nanorods is shown in Figure 17. The wave numbers ($cm^{-1}$) of dominant peaks obtained from absorption spectrum are presented in Table 17. The absorption peak at 462.86 $cm^{-1}$ indicates the presence of lead [55]. The observed wave numbers for amines, amides and amino acids indicate the presence of protein [56]. The details are presented in Table 18. The amino group is one of the key factors in controlling the Pb nanoparicles.

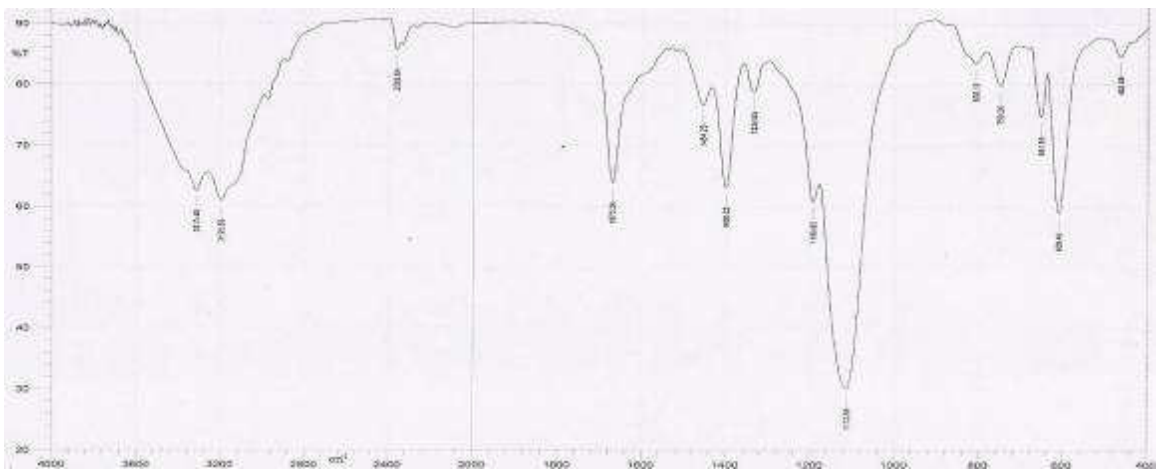

**Fig.17.** FT-IR Spectrum of Lead Nanorods

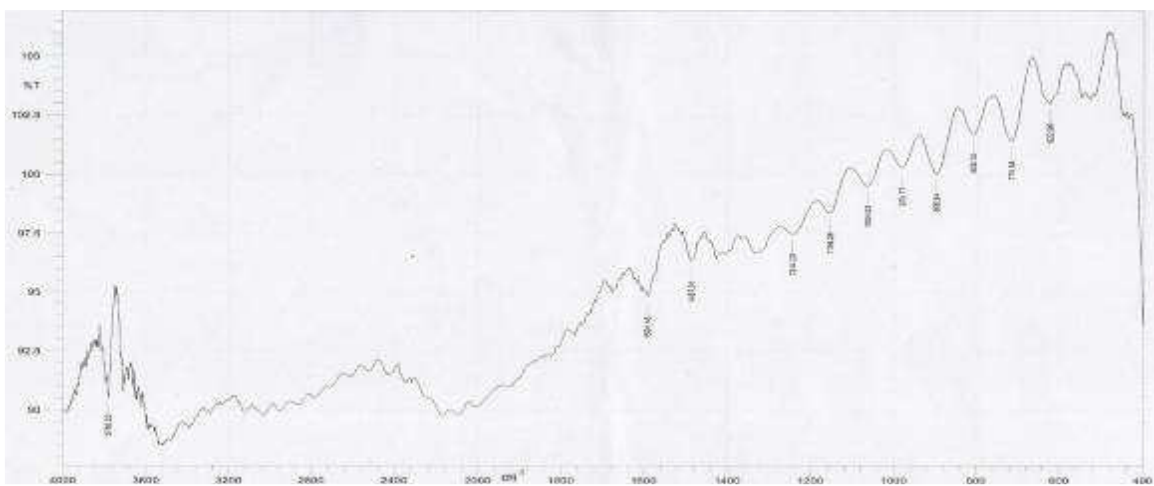

**Fig.18.** FT-IR Spectrum of Bulk Lead

The protein material are encapping the metal particles and likely to serve as a capping / stabilizing agent [57–59]. It is concluded from earlier report NH /C-O groups bind with metals (lead) nanoparticles [60]. Amino acid residues and peptides of proteins have the stronger ability to bind metal and very high affinity to bind with metals. So that protein itself can act as an encapsulating agent and possibly form a layer covering the metal nanoparticles. This layer prevents agglomeration, protects the nanoparticles from agglomerization and thereby stabilizes the medium. [61, 62].

The observed bands at wave numbers 3313 cm$^{-1}$, 3195 cm$^{-1}$, 2362 cm$^{-1}$, 1670 cm$^{-1}$, 1454 cm$^{-1}$, 1400 cm$^{-1}$, 1334 cm$^{-1}$ and 1193 cm$^{-1}$ imply the presence of protein/peptide on the nanoparticle surface. The NH/C-O groups also indicate the presence of bio-molecules like nitrates and Carboxylic Acids. C–O–C stretching of wave numbers 1193 cm$^{-1}$, 808 cm$^{-1}$ and 750 cm$^{-1}$ indicates polysaccharides. It is noted here that Glucomannan is a water-soluble polysaccharide (hemicelluloses) - straight-chain polymer - a super absorbent is present in konjac. Oxidations of the synthesized lead nanoparticles are prevented by polysaccharides which are serving as reducing agent. Chealating compounds sediment the lead nanoparticles from residues.

**Table.17.** FTIR Functional groups analyses of Lead Nanorods-konjac

| Vibrational Assignment / Functional Groups | Observed Wave Number (cm$^{-1}$) | Visible Intensity |
|---|---|---|
| C-H γ + OH γ + N-H γ$_{as}$ | 3313.48 | Broad |
| N-H γ | 3195.93 | Broad |
| N-H γ [-SO$_2$ H-OH γ] | 2362.64 | W |
| C=C γ + C-H γ + C=O γ + N-O γ + C=N γ | 1670.24 | S |
| CH$_3$ δ$_{as}$ + CH$_2$ δ$_s$ + N=O γ + NH α | 1454.23 | VW |
| CH$_2$ δ + OH α + C-O γ + C-CHO (skeletal) + COO$^-$ γ$_s$ + C-N γ + N-H α + CO$_2^-$ γ$_s$ + N=O γ + C-F γ | 1400.22 | S |
| C-O γ + OH α + C-CHO (skeletal) + COO$^-$ γ$_s$ + C-N γ + NO$_2$ γ$_s$ + N$_3$ γ$_s$ + SO$_2$ γ$_{as}$ + C=S γ | 1334.65 | W |
| C-CHO (skeletal) + C-O-C γ + NH$_3^+$ α + N=O γ + C-F γ | 1193.85 | W |
| C-O-C γ$_{as}$ + C-CHO (skeletal) + C-O γ + N-N γ + C-F γ | 1112.85 | VS |
| C-H β + C-O-C γ + N-H ω + N-H τ + O-N γ | 808.12 | VW |
| C-C β (ring puckering) + C-H β + O-H β + C-O-C γ + N-H ω + N-H τ + O-N γ + C-S γ + C-Cl γ + C-Br γ | 750.26 | W |
| O-N=O δ + C-S γ + S-O γ + C-Br γ | 651.89 | W |
| OCN(deformation) + C-S γ + S-O γ + C-I γ | 609.46 | S |
| S-S γ | 462.86 | VW |
| Abbreviations: α -in-plane bending; β -out-of plane bending; ω –wagging; τ -twisting; γ -stretching; δ – bending; γ$_s$ - symmetric stretching; γ$_{as}$ - asymmetric stretching; δ$_s$ -symmetric bending; δ$_{as}$- asymmetric bending; | | |
| S – Strong; W – Weak; VS - Very strong; VW - Very Weak; | | |

Phenolic compounds consisting of a hydroxyl group (OH) bonded directly to an aromatic hydrocarbon group. Presence of OH group (Wave numbers 3313$^{-1}$ and 750 cm$^{-1}$) along with aromatic compounds indicates phenolic compounds. C-O stretching vibration indicates the secondary alcohol. [63]. Presence of alcohol is observed from the wavenumbers 3313 cm$^{-1}$, 1670 cm$^{-1}$, 1400 cm$^{-1}$, 1334 cm$^{-1}$ and 750 cm$^{-1}$ due to C-O and O-H groups. It is known that Pb nanoparticles are highly reactive and transfer electron to N$_2$O, O$_2$, etc. The synthesized lead nanoparticles, while on transferring electrons to air, photo-initiators like phenolic compounds and alcohols play some roles to prevent such electron transfer. They absorb light and form hydrated electrons to reduce metal ions to metal. It is noted here that the flying powder of konjac is a polyol (natural polymer) which precipitates soluble heavy metal ions in water.

wavenumbers 3313 cm$^{-1}$, 1670 cm$^{-1}$, 808 cm$^{-1}$ and 750 cm$^{-1}$ indicate the flavanoids which are anti-oxidant substances possess reducing power and can donate electrons. They prevent the oxidation of lead nanoparticles. Proteins in the konjac aqueous extract bind to lead nanoparticles through amine and carboxylates (COO $^{-}$ at the wave number 1400 cm$^{-1}$ and 1334 cm$^{-1}$).

Water has been used as solvent in the synthesis process. So that the hydroxyl group (OH) is occurred predominantly (Wavenumbers 3313 cm$^{-1}$, 1400 cm$^{-1}$, 1334 cm$^{-1}$ and 750 cm$^{-1}$) which indicates the moisture surroundings of nanoparticles. Br-stretching (Wave numbers 750 cm$^{-1}$ and 651 cm$^{-1}$) indicates that mixing of KBr with lead nanoparticles for making pellet while FTIR analyses.

**Table.18.** FTIR Bio – Chemical Compounds analyses

| Bio – Chemical Compounds | | Wave Number (cm$^{-1}$) |
|---|---|---|
| **Amines** | N–H stretching | 3313, 3195, 2362 |
| | N–H bending | 1454, 1400, 1193 |
| | C–N stretching | 1670, 1400, 1334, |
| **Amides** | N–H stretching | 3313, 3195, 2362 |
| | C–O stretching | 1670, 1400, 1334 |
| **Amino acids** | N–H stretching | 3313, 3195, 2362 |
| | N–H bending | 1454, 1400, 1193 |
| | C–O stretching | 1670, 1400, 1334 |
| **Nitrates** | N–H bending | 1454, 1400, 1193 |
| **Carboxylic Acids** | O–H stretching | 3313, |
| | C- O stretching | 1670, 1400, 1334 |
| **Polysaccharides** | C–O–C stretching | 1193, 808, 750 |
| **Chelate Compounds** | O–H stretching | 3313 |
| **Aromatics / Flavanoids** | C=H stretching | 3313 |
| | C=C stretching | 1670 |
| | C–H out of plane bending | 808, 750 |
| **Alcohols** | C–O stretching | 1670, 1400, 1334 |
| | O–H group | 3313, 1400, 1334, 750 |
| **Phenols** | O–H group | 3313, 750 |

For a nano size grain, the atomic arrangements on the boundaries differ greatly from that of bulk crystals: both in coordination number and bond lengths, showing some extent of disorder. Crystal symmetry is thus degraded in nano size grains. The degradation in crystal symmetry results in the shifting of IR active modes. If grain size of the sample is in nano size, all the IR active modes will be shifted to the high frequency side / low wavelength. The FT-IR analysis of bulk lead has also been done and the FT-IR spectrum has shown in Figure 18. All the IR active modes of bulk lead have been compared with IR modes in the spectrum of Pb nanoparticles to demonstrate the shifting of IR modes towards high frequency side / low wave length side. We did not observe such shifting of IR modes towards high frequency side / low wave length side after reducing the size of bulk lead to nanoparticles.

### 3.3. Anti-Bacterial studies of Spherical Pb nanoparticles

Nanomaterials are the leading requirement of the rapidly developing field of nanomedicine, bionanotechnology. Nanoparticles usually have better or different qualities than the bulk material of the same element and have immense surface area relative to volume. For centuries, People have avoided lead, in their normal life, unlike other metals due its cumulative poison nature. From the preliminary screening anti-bacterial study against *E.coli* bacteria, it is observed that lead nanoparticles act different from bulk lead, like an inert material and show mild activity when using large quantity but, how this material will act in other biological cells like plants and animals which should be further investigated elaborately. Moreover, bulk lead dissolves in acetic acid, nitric acid and sodium hydroxide solution but these materials unable to dissolution of lead nanoparticles.

Antibacterial activities of lead nanoparticles were evaluated by cup and plate method. A concentration of 50 mcg / ml sample solution was used. Zone of Inhibition (ZOI) was measured from this microbiology assay. From the result, it is observed that the sample possessed very mild or no activity against *E.coli* and showed diameter of zone of inhibition against 5 mm (Figures 19 and 20). The sample shows no antimicrobial activity against E.coli when examined using cup and plate method. Mild zone of inhibition of 0.5 cm was found when the sample of 20mg was directly kept on the surface of the plate with E.coli.

In a solid material, the surface-area-to-volume ratio (SA: V) or Specific Surface Area (SSA) is an important factor for the reactivity that is, the rate at which the chemical reaction will proceed. Materials with large SA: V (very small diameter) reacts at much faster rates than monolithic materials, because more surfaces are available to react.

For studying, changes in Specific Surface Area (SSA) of nanoparticles and its effects on antibacterial activities nanoparticles, we have compared SSA of lead nanoparticles and silver Nanoparticles of our earlier study [46]. The details are presented in Table 19. Our earlier study suggests that that increased SSA results enhance the antibacterial activities of Silver nanoparticles. From this comparative study, though lead nanoparticles have SSA more than silver nanoparticles, the antibacterial activities of lead nanoparticles is less than Silver nanoparticles. It indicates that the reactions of nanoparticles with bacterial cell wall / fluid are also the determining factor of antibacterial activities.

**Table.19.** Comparison of Surface Area to Volume Ratio and antibacterial activities of Lead Nanoparticles on Escherichia Coli (Gram Negative bacteria)

| Nano particles | Particle Size (nm) | Surface Area ($nm^2$) | Volume ($nm^3$) | Surface Area to Volume Ratio | SSA ($m^2g^{-1}$) | Diameter Inhibition Zone (mm) | *E.coli* Bacteria SSA ($m^2g^{-1}$) |
|---|---|---|---|---|---|---|---|
| Lead | 10 | 314.15 | 523.59 | 0.53 | 52.35 | 5 | 20.09 |
| Silver | 24 | 1809 | 7235 | 0.25 | 24 | 12 | |

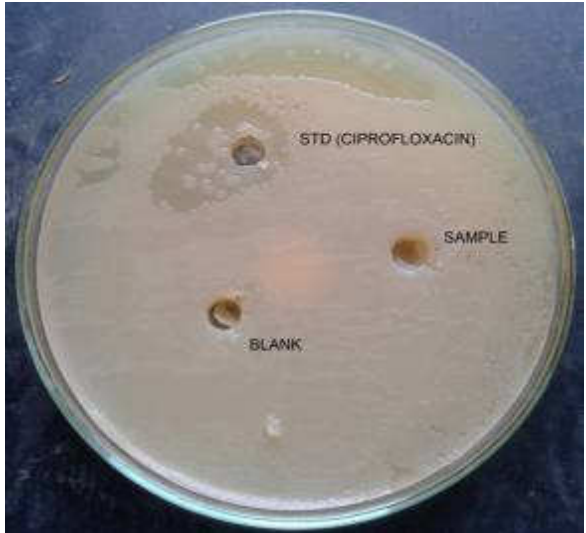 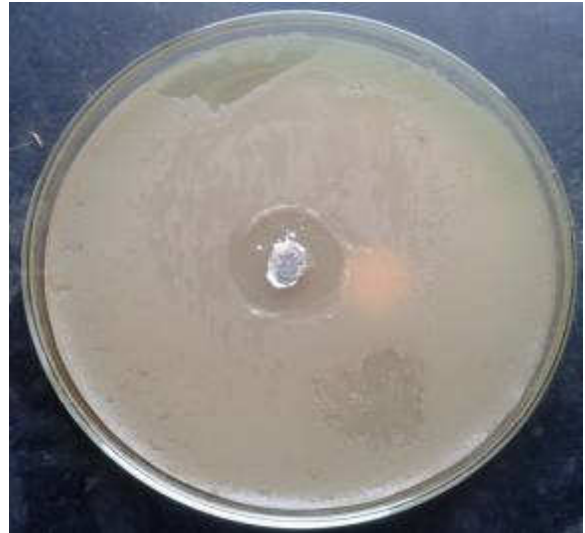

**Fig.19.** No Zone of inhibition against *Escherichia coli* bacteria

**Fig..20**. Zone of inhibition diameter against *Escherichia coli* 5mm.

The SSA of cells has an enormous impact on their biology. SSA places a maximum limit on the size of a cell. An increased SSA also means increased exposure to the environment. Greater SSA allows more of the surrounding water to be shifted for nutrients. Increased SSA can also lead to biological problems. More contact with the environment through the surface of a cell increases loss of water and dissolved substances. High SSA also present problems of temperature control in unfavorable environments.

SSA affects the rate at which particles can enter and exit the cell whereas the volume affects the rate at which material are made or used within the cell. These substances must diffuse between the organism and the surroundings. The rate at which a substance can diffuse is given by Fick's law.

$$Rate\ of\ Diffusion \propto \frac{surface\ area \times concentration\ difference}{distance} \quad \ldots\ldots\ldots\ldots (20)$$

So rate of exchange of substances depends on the organism's surface area that's in contact with the surroundings. Requirements for materials depend on the volume of the organism, so the ability to meet the requirements depends on the SSA.

We have made an attempt to study the SSA of bacteria and its reactivity to antibacterial activities of nanoparticles. For this study, we have compared SSA of *E.coli* with nanoparticles of lead and silver. *E.coli* details (Cell length: 2 μm or $2 \times 10^{-6}$ m, diameter: 0.8 μm or $0.8 \times 10^{-6}$ m, total volume: $1 \times 10^{-18}$ m$^3$, surface area: $6 \times 10^{-12}$ m$^2$, wet weight: $1 \times 10^{-12}$ g, dry weight: $3.0 \times 10^{-13}$ g) has been extracted from The CyberCell Database- CCDB and SSA calculated accordingly [64].

Bacteria, viruses and fungi all depend on an enzyme to metabolize oxygen to live. Silver interferes with the effectiveness of the enzyme and disables the uptake of oxygen, thereby killing the microbes but in the case of lead, it does not interfere with the enzyme or showing less interference at high concentration. It is observed from our earlier study of silver nanoparticles that SSA of both nanoparticles and bacteria are playing major roles in their inter reactions and influencing and affecting factors of ZOI. The present study reveals that apart from the SSA of them, the interferences, bio-chemical reactions and affinity of metal nanoparticles with cell wall / fluid of microbes is also one of the main factors.

Exposure of bacteria cell to the environment and rate of exchange of substances that is in contact with its surroundings depends on the SSA of bacteria. E.coli shows same exposure to the environment / surroundings in which lead and silver nanoparticles are exist. The SSA of lead nanoparticles is more than silver nanoparticles. It means lead nanoparticles have more reactive sites than silver nanoparticles. Also, due to its lesser size, it can easily enter into the E.coi cell more than silver nanoparticles while E.coli exchanging the surroundings substances.

These factors are supporting for the reactions between lead nanoparticles and E.coli but there is no bio-chemical reactions between them or very less reactions while on using more concentration of sample and it is opposite, in the case of silver nanoparticles. The more bio-chemical reactions / interferences between silver nanoparticles and E.coli produce unfavourable surroundings results in more damages to bacteria and increased Zone of Inhibition.

## 4. Conclusions

We have synthesized rod and spherical shaped Pb nanopowder by electrolysis using a bioactive compound - konjac aqueous extract and have analyzed their characters successfully by various analysis tools. The synthesized lead powder has potential applications and can be used in many diverse industries such as scientific, nuclear Oil & gas exploration, anti-corrosive paints, lubricating, and sinter materials, catalysts, explosives industries and also in medical, electrical fields. This synthesis method is a unique, novel and double-step procedure with good reproducibility and has so far not been used for nanoparticles preparation.

XRD confirms that the prepared nanoparticles are lead which agrees with EDS analysis. The approximate size calculated from XRD is 8 nm which corroborates with the size 10 nm assessed from TEM and the size 55 nm is observed from Particle Size Analyzer. The spherical shape is observed from TEM. The result of SEM gives a conclusion about the rod shape of the synthesized nanoparticles. The melting point observed from DSC is well agreement with the lead nanoparticles synthesized by other method and sharper melting peak show the purity as well as the smaller size of the synthesized sample. AAS result concludes that absorbance of the sample is equal to the absorbance of the standard lead solution. FTIR confirms binding of bio-molecules with lead nanoparticles. Anti-bacterial assay of lead nanoparticles against E.coli shows that lead nanoparticles act like an inert material and show mild activity when using large quantity due to very less bio-chemical reactions in between them.

The observed results conclude that the oxidation of Pb nanoparticles is prevented by the stabilizing ability of the konjac aqueous extract. Generally, metal nanoparticles strongly absorb light in the visible region due to Surface Plasmon Resonance (SPR). Likewise, light is absorbed by Pb nanoparticles and the absorbed light promotes the reactions / effects of the sunlight on Pb nanoparticles. On this basis, an attempt has been made to find the effects of the sunlight on nanomaterials which concludes that the sunlight can be utilized for dual function i.e. as a dryer for the synthesized nanomaterials and as a morphological changer. It also explicates that it is possible to change the shape of Pb nanoparticles (konjac extract added) from spherical to rod by the influence of sunbeams, even after completion of the entire synthesis process. The research work is under process to find the possibility of morphological effect changes by sunlight, in other metal nanoparticles with konjac extract or other bio-molecules.

## ACKNOWLEDGEMENTS

The authors express their immense thanks to Central Electrochemical Research Institute, (CECRI, Karaikudi), Karunya University (Coimbatore), Indra Gandhi Centre for Atomic Research (IGCAR, Kalpakkam), The Standaard Fireworks Rajaratnam College for Women (Sivakasi), Ayya Nadar Janaki Ammal College (ANJAC, Sivakasi) for providing laboratory instruments to analyze the samples, Dr.S.Amirthapandian of IGCAR for his assistances and valuable guidances in TEM analyses and to Dr.M.Palanivelu, of Arulmigu Kalasalingam College of Pharmacy (Kalasalingam University, Krishnankoil, India) for assistances in anti-bacterial studies. They also acknowledge assistances and encouragements of staff & management of PACR Polytechnic College, Rajapalayam, India and ANJAC (Sivakasi, India).